\documentclass[iop]{emulateapj}
\usepackage{newtxtext,newtxmath}
\usepackage[usenames]{color}
\usepackage{hyperref}

\hypersetup{colorlinks=true,linkcolor=black,citecolor=blue,filecolor=cyan,urlcolor=magenta}

\newcommand{\ebv}{$E_{B-V}$}

\newcommand{\FUVmV}{${\it FUV}\!-\!V$}

\newcommand{\ZH}{[$Z$/H]}
\newcommand{\aFe}{[$\alpha$/Fe]}
\newcommand{\kms}{km\,s$^{-1}$}

\newcommand{\plm}{$\,\pm\,$}
\newcommand{\Rgal}{R_{\rm gal}}
\newcommand{\cM}{{\cal{M}}}

\newcommand{\mybibitem}[3]{\bibitem[{#1}({#2})]{#3}}
\newcommand{\mybibthree}[4]{\bibitem[{#2}({#3}){#1}]{#4}}
\submitted{Received 2018 February 8; revised 2018 February 28;
  accepted 2018 March 4; published 2018 April 9}
\journalinfo{Published in The Astrophysical Journal, {\bf 857}, 16 (2018)}

\shorttitle{The UV Upturn: caused by dissolved metal-rich GCs?}
\shortauthors{Paul Goudfrooij}

\begin{document}

\title{Dissolved Massive Metal-Rich Globular Clusters can cause the Range of UV
  Upturn Strengths found among Early-Type Galaxies} 
\definecolor{MyBlue}{rgb}{0.3,0.3,1.0}
\author{Paul Goudfrooij} 
\affil{Space Telescope Science Institute, 3700 San Martin Drive,
  Baltimore, MD 21218, USA;
  \href{mailto:goudfroo@stsci.edu}{\color{MyBlue}goudfroo@stsci.edu}}  



\begin{abstract}
I discuss a scenario in which the ultraviolet (UV) upturn of giant early-type
galaxies (ETGs) is primarily due to helium-rich stellar populations that formed
in massive metal-rich globular clusters (GCs) which subsequently dissolved in
the strong tidal field in the central regions of the massive host galaxy. 
These massive GCs are assumed to show UV upturns similar to those observed
recently in M87, the central giant elliptical galaxy in the Virgo cluster of
galaxies. Data taken from the literature reveals a strong correlation between
the strength of the UV upturn and the specific frequency of metal-rich GCs in
ETGs. 
Adopting a Schechter function parametrization of GC mass functions, simulations
of long-term dynamical evolution of GC systems show that the observed
correlation between UV upturn strength and GC specific frequency can be 
explained by variations in the characteristic truncation mass $\cM_{\rm c}$ such
that $\cM_{\rm c}$ increases with ETG luminosity in a way that is
consistent with observed GC luminosity functions in ETGs. 
These findings suggest that the nature of the UV upturn in ETGs and the
variation of its strength among ETGs are causally related to that of
helium-rich populations in massive GCs, rather than intrinsic properties of
field stars in massive galactic spheroids.
With this in mind, I predict that future studies will find that [N/Fe]
decreases with increasing galactocentric radius in massive ETGs, and that such
gradients have the largest amplitudes in ETGs with the strongest UV upturns.   
\end{abstract}




\section{Introduction} 
\label{s:intro}

Starting in the 1970's, various ultraviolet (UV) space observatory missions have
firmly established that luminous early-type galaxies (hereafter ETGs) and bulges of
spiral galaxies feature far-UV (FUV) emission whose strength rises shortwards
of $\sim$\,2000 \AA\ 
\citep[e.g.,][]{codwel79,bert80,burs+88,ocon+92,dorm+95,brow+97}. This
phenomenon is widely known as the ``UV upturn,'' which was originally
unexpected because ETGs were generally believed to be old ``red and dead''
stellar populations.

Studies of the UV-to-optical spectral energy distribution of ETGs showed that
the UV upturn is well-fit by a narrow range of effective temperatures between
$\sim$\,20,000 and 25,000 K \citep[e.g.,][]{brow+97}. It is commonly believed
that the UV upturn in ETGs is mainly produced by extreme horizontal branch
(EHB) stars, also known as hot subdwarfs, and their progeny (see, e.g., the
reviews by \citealt{ocon99} and \citealt{yi08}).  
These are helium-core-burning stars with extremely thin hydrogen
envelopes. Various scenarios for their formation have been proposed:\ strong
mass loss on the red giant branch (RGB; this likely mainly occurs at high
metallicity, see \citealt{greren90}), high helium abundance
\citep[e.g.,][]{dorm+95,yi+97}, and/or mass transfer of RGB envelopes in binary
systems \citep{han+07}.   

Important constraints to the nature of the UV upturn in nearby ETGs were
introduced by \citet{burs+88} who showed that the $m(1550)-V$ color (hereafter
\FUVmV) anticorrelates strongly with both velocity dispersion ($\sigma$) and
the Lick Mg$_2$ absorption-line index. Since the latter was thought to be
mainly a metallicity indicator and because $\sigma$ also correlates strongly
with Mg$_2$, \citeauthor{burs+88} argued that stellar metallicity is the
fundamental parameter underlying these relations.

Another key feature of the UV upturn in ETGs is that its spatial distribution
is more centrally concentrated than the optical light \citep{ohl+98,cart+11}.
As such, matching of measurement apertures between the FUV and other
wavelengths is important in the interpretation of correlation studies. This
point proved to be relevant in understanding apparent inconsistencies between
different studies of the \FUVmV\ versus Mg$_2$ relation using the \emph{GAlaxy
  Evolution eXplorer} \citep[\emph{GALEX};][]{mart+05}: \citet{rich+05} used a
sample of ETGs covering a significant range of distances ($z < 0.2$) along with
fixed measurement apertures and did \emph{not} recover the \citet{burs+88}
relation. However, several other studies used samples of ETGs for which UV and
optical measurement apertures were matched on a galaxy-by-galaxy basis
\citep{bure+11,jeon+12} or samples of cluster galaxies to eliminate the
distance effect \citep{bose+05,smit+12a}. The latter studies all do recover the
\citet{burs+88} relation (some using the Mg$b$ index rather than Mg$_2$). Other
relations between UV upturn and ETG parameters that were established by these
studies were anticorrelations of \FUVmV\ with $\sigma$, age, total metallicity
\ZH\, and $\alpha$-element abundance enhancement \aFe.
These anticorrelations are not expected in the scenario where hot
subdwarfs are produced by strong mass transfer of RGB envelopes in binaries,
in which the dependence of \FUVmV\ on age and metallicity is insignificant  
\citep{han+07}. Consequently, many recent studies of the UV upturn
have concentrated on the single-star mechanisms (but see
Section~\ref{sub:ener}.1).   

A pivotal discovery in the context of the nature of the UV upturn was that of
massive FUV-bright globular clusters (GCs) in M87, the central galaxy of the
Virgo cluster\footnote{To date, M87 is the only ETG with a UV
  upturn for which FUV photometry of GCs has been published.}
(\citealt{sohn+06}; see also \citealt{peac+17}). Many of those 
GCs in M87 were found to feature \FUVmV\ colors similar to, or even bluer than,
ETGs with the strongest UV upturns. Interestingly, several massive
\emph{metal-rich} GCs in M87 have both \FUVmV\ and ${\it NUV}\!-\!V$
colors consistent with those of massive ETGs\footnote{In contrast, the
  metal-poor GCs in M87 generally feature bluer UV\,--\,optical colors 
  than those of ETGs with UV upturns, especially in ${\it
    NUV}\!-\!V$, which is likely due to metallicity-dependent line blanketing
  in the NUV \citep[e.g.,][]{dorm+95}.}. Among metal-rich GCs in our Galaxy,
this behavior is only seen in NGC~6388 and NGC~6441, two massive
GCs with very hot EHBs \citep[see][]{rich+97}, which are
thought to be due to a high He abundance in a significant fraction of their
constituent stars \citep{caldan07,tail+17}.  

In fact, a myriad of recent studies of multiple stellar populations
in GCs in the Local Group revealed that light-element
abundance variations within massive GCs are the norm rather than the
exception \citep[see, e.g., the review by][]{grat+12}.
The strongest abundance patterns emerging from spectroscopic studies of GCs are
that [Na/Fe] correlates with [N/Fe] and generally anticorrelates with [O/Fe]
and [C/Fe] \citep[e.g.,][]{sned+92,roed+14}. These patterns are thought
to be the result of proton-capture reactions at $T \ga 2\times10^7$ K such as
the CNO and NeNa cycles \citep[e.g.,][]{grat+12}. 
The material responsible for these abundance variations is generally thought to
originate in winds of stars massive enough to host such high temperatures in
their interiors (``polluters''). Subsequent episodes of star formation in GCs
with masses and escape velocities high enough to retain these winds may have
caused the abundance variations seen today within such GCs
\citep[see, e.g.,][]{derc+08,goud+11a,conr12,renz+15}.  

Recent studies have shown that GC mass is indeed an important parameter
in the context of light-element abundance variations in GCs. Its relevance was 
first suggested by the results of \citet{carr+10}, who showed that the extent of
the Na--O anticorrelation within GCs scales with GC mass. More recently,
\citet{milo+17} studied high-precision \emph{Hubble Space Telescope (HST)}
photometry of RGB stars in 57 Galactic GCs using a selection 
of filters that emphasizes abundance spreads in N and O. They found a strong
correlation between the fraction of stars enhanced in [N/Fe] and the GC
mass. Furthermore, studies using integrated-light spectroscopy of GCs in M31
and our Galaxy show a similarly strong correlation between overall [N/Fe] and GC 
mass (\citealt{schi+13}; T.\ H.\ Puzia \& P.\ Goudfrooij 2018, in preparation). 

Importantly with regard to the UV upturn, the proton-capture 
processes at high temperatures that produce enhancements in [N/Fe] and [Na/Fe]
also produce He enhancement, one of the main contenders for producing the hot
EHB stars in GCs, including those in M87 \citep[see][]{kavi+07}. While direct 
measurements of the He abundance (hereafter $Y$) are generally difficult
to obtain for stellar populations, it is well known that increases in $Y$ cause
hotter and bluer main sequences in color-magnitude diagrams
\citep[e.g.,][]{dott+07,piot+07,cass+17}. 
This effect was used by \citet{milo15} to show that the spread in $Y$ found
within Galactic GCs is strongly correlated with GC mass, similar to the case of
[N/Fe] spread mentioned above. 
This correlation is consistent with recent photometric studies of HB stars in
GCs using \emph{HST}, which clearly indicate a need for significant He
enhancement to fit the multi-color photometry of the HBs in the most massive
GCs and to explain the finding that the hottest types of HB stars (``blue-hook
stars'') exist only in the most massive GCs \citep{brow+10,brow+16}. 

In this paper, we investigate the idea that there is a physical connection
between the UV upturn in ETGs and the He-enhanced populations in massive
GCs. This connection was suggested earlier:\ 
\citet{chun+11} constructed simple stellar population (SSP) models
including effects of He enhancement and showed that the UV upturns of
ETGs are well fit by their models,
while \citet{bekk12} studied this connection using numerical simulations. 
Here, we explore the possibility that the range of UV upturn strengths found
among ETGs is caused by He-rich stars formed in massive GCs that subsequently
disrupted in the strong tidal field of the inner regions of their host galaxies.  

\section{GC Systems in Early-Type Galaxies}
\label{s:GCSes}

While GCs typically constitute a very small fraction of the stellar mass in
``normal'' galaxies, their properties contain important clues to the assembly
histories of their host galaxies. Infrared studies of star formation within
molecular clouds have shown that stars typically form in a clustered fashion
\citep[bound clusters or unbound associations;][]{ladlad03,port+10}. Most star
clusters with initial masses $\cM_0 \la 10^4\;M_{\odot}$ are thought to
evaporate into the field population of galaxies within a few Gyr, but the
currently surviving massive GCs represent important probes of the chemical
evolution occurring during the main epochs of star formation within a galaxy's
assembly history.  

One of the most studied observational parameters in the context of the
GC--galaxy connection is the \emph{specific frequency} of GCs (hereafter
$S_N$). This parameter was introduced by \citet{harvdb81}, who defined it as the
number of GCs per unit galaxy luminosity in units of $M_V = -15$.
$S_N$ essentially measures the number of GCs that survived
dynamical evolution over the galaxy's lifespan relative to the total luminosity
of stars that evaporated \emph{out of} star clusters or associations during that
period. As such, its value is determined by the shape of the initial cluster
mass function (ICMF) and the time-dependent effects of dynamical evolution of
GCs.\footnote{This is discussed further in Sections~\ref{sub:GCdiss} and
  \ref{sub:ener}.}    

Several studies have established that $S_N$ varies greatly among galaxies. The
general trend among massive galaxies is that $S_N$ increases significantly with
increasing galaxy luminosity: $S_N$ for the most luminous ETGs (with
$M_V \sim -23$) is a factor of $\sim$\,5 higher on average than that
for ETGs with $M_V \sim -20$ \citep{peng+08}. Conversely, the trend is
opposite among galaxies fainter than $M_V \sim -20$, for which $S_N$
increases with \emph{decreasing} galaxy luminosity, with several faint
dwarf galaxies with $-16 \la M_V \la -12$ showing $S_N$ values larger
than those of the most luminous ETGs
\citep[e.g.,][]{forb05,millot07,peng+08,geor+10,harr+13}.  

Deep imaging studies with the \emph{HST} and large ground-based telescopes have
shown that ``normal'' luminous ETGs typically contain rich GC systems  
featuring wide, bimodal optical color distributions
\citep[e.g.,][]{brostr06,peng+06}. Spectroscopy showed that such GCs are nearly
universally old ($\ga 10$ Gyr), independent of their color
\citep[e.g.,][]{puzi+05,brod+12}, implying that their colors mainly indicate
differences in metallicity. The ``red'' GCs feature colors and spatial
distributions that generally follow those of the underlying 
spheroid light (i.e., the ``bulge''), while the ``blue'' GC subsystems typically
show much more extended (``halo-like'') spatial distributions, consistent with
those of metal-poor GCs in our Galaxy and M31 
\citep[e.g.,][]{geis+96,rhozep04,brostr06,harr+10,puzi+14}.
The overall number ratio $N$(red)/$N$(blue) changes with ETG mass or
luminosity. Low-mass ETGs with $M_V \ga -18$ host virtually only blue GCs. 
For more massive ETGs, the fraction of red GCs increases with galaxy luminosity
up to $M_V \approx -22$, beyond which it decreases again due to large numbers
of blue GCs, mainly located in the outer regions of the most massive ETGs
\citep{peng+08}.  

The general picture that has emerged from these studies is that the blue GCs
are likely to represent remnants of the earliest stages of hierarchical merging
of gas-rich, metal-poor protogalactic dwarfs, which caused high GC formation
rates in regions with high star formation rate surface density
\citep[e.g.,][]{burg+01,moor+06,peng+08}, while the red GCs were formed during 
mergers of more massive gas-rich galaxies at high redshift
\citep[e.g.,][]{peng+08,krui15}, perhaps situated within more massive halos that were
able to retain the chemically enriched gas outflows from earlier star formation
episodes. 
The large numbers of blue GCs found in the outskirts of the most massive ETGs
likely reflect relatively large numbers of accreted dwarf galaxies, while the
increase in $S_N$ of red GCs with increasing ETG luminosity is thought to
indicate an increasing mass fraction of higher-mass protogalaxies that
underwent starbursts at high SFR surface density
\citep{ashzep92,peng+08,krui15}.

\section{Relations between UV Upturn Strength and Population and GC System
  Properties}  
\label{s:rela} 

To explore the relevance of the specific frequency of GCs in the production of
the UV upturn in ETGs, we select two samples of nearby ETGs based on their source of
UV data (IUE vs.\ GALEX) and the availability of dynamical and stellar
population properties as well as GC specific frequency data for both
metal-poor and metal-rich GC subpopulations in the literature.

The first sample is a subset of that studied by \citet{dorm+95}, using the
aperture-matched IUE, $V$-band, and Mg$_2$ data from \citet{burs+88}. We remove
galaxies with strong dust extinction in their inner regions, since this renders a
significant uncertainty to their intrinsic \FUVmV\ colors. This eliminates the
galaxies NGC 1052, NGC 2768, NGC 4111, NGC 4125, NGC 4278, NGC 4374, and NGC
5846 \citep{hans+85,goud+94b,vanfra95,goutri98,mart+04,laue+05,mase+11}. The
resulting sample is henceforth referred to as the ``IUE sample.'' 
Stellar population parameters for eight galaxies in the IUE sample are taken
from \citet{thom+05}, using spectroscopic measurements of indices H$\beta$,
Mg$b$, Fe5270, and Fe5335 in the Lick system 
within an aperture of $R_{\rm eff}/10$, where $R_{\rm eff}$ is the effective
radius of the galaxy. The ages, metallicities (\ZH) and \aFe\ ratios
are determined using the SSP models of \citet{thom+03}.  For the nine
other galaxies in the IUE sample, we use Lick-system line index measurements
from other studies (see Table~\ref{t:IUEsample}) and use the \citet{thom+03} SSP
models to place their SSP-equivalent ages, \ZH, and \aFe\ measurements in the
same system as that of \citet{thom+05}, using linear interpolation between the
model grid points.  In this context we use the diagram of
$\left< \rm{Fe} \right>$ (= $0.5 \times {\rm (Fe5270+Fe5335)}$) versus Mg$b$ to
provide a first estimate of the \aFe\ ratio, and the [MgFe]$'$  versus H$\beta$
diagram to estimate the ages and \ZH\ values.\footnote{The [MgFe]$'$ index was
  chosen because of its independence of variations in [$\alpha$/Fe] ratio and
  IMF slope \citep{thom+03,vazd+10}.} 
The three population parameters are then modified using a small grid around the
first estimates, and the best error-weighted fit to H$\beta$, Mg$b$, and
$\left< \rm{Fe} \right>$ is selected.  

The second sample of galaxies is taken from the SAURON project
\citep[e.g.,][]{deze+02}, for which \emph{GALEX} images and optical
spectroscopy were analyzed by \citet{bure+11} and \citet{jeon+12}, using
measurements in an aperture of $R_{\rm e}/2$ for each galaxy. 

For both galaxy samples, specific frequencies $S_N$ of metal-rich (red) GCs are 
taken from recent papers in the literature that used high-quality GC colors
from data taken with the Advanced Camera for Surveys (ACS) aboard
\emph{HST}. References are given in Tables~\ref{t:IUEsample} and
\ref{t:SAURONsample} for the IUE and SAURON samples, respectively.
For the Fornax ellipticals NGC 1399 and NGC 1404 in the IUE sample, we
determine metal-rich GC fractions from the GC photometry tables of the ACS
Fornax Cluster Survey \citep[see][]{jord+15}. To stay consistent with the
results for Virgo galaxies by the ACS Virgo Cluster Survey, we adopt the
procedures and GC selection criteria of \citet{peng+06} in this
context. Briefly, to establish red GC fractions
(designated $f_{\rm  red}$ in Tables~\ref{t:IUEsample} and
\ref{t:SAURONsample}), we apply the Kaye's Mixture Model
\citep[KMM,][]{mclbas88,ashm+94} to fit two Gaussians to the GC color
distribution using the homoscedastic case, where $\sigma$ is the same for both
Gaussians.  
Specific frequencies of red GCs were determined following
\citet{peng+08}, normalizing the derived number of such GCs in the galaxy by the
galaxy luminosity in the SDSS (Sloan Digital Sky Survey) $z$ passband (similar
to the F850LP passband of \emph{HST}/ACS):  
\begin{equation}
  S_{N,\,z,\,{\rm red}} = N_{\rm GC,\,red} \times 10^{0.4(M_z + 15)}
\label{eq:Sn_red}
\end{equation}
For galaxies in the ACS Virgo Cluster Survey, we adopt the total $z$-band
magnitudes in the AB system from \citet{peng+08}. For other galaxies with
$S_N$ data that were derived from total $V$-band luminosities in the
literature, integrated $V\!-\!z_{\rm AB}$ colors are determined in two different
ways. For NGC 4889, we simply use the observed integrated $V\!-\!z$ color
from \citet{eise+07}, assuming $z_{\rm AB} = z_{\rm Vega} + 0.518$. For the
other galaxies, $z$-band photometry is not available and we use high-quality
integrated colors in other passbands along with the SSP models of
\citet[][hereafter BC03]{bc03} to estimate $V\!-\!z$ as follows. We use
\href{http://ned.ipac.caltech.edu}{NED} 
to obtain the integrated colors $(V\!-\!R)_J$ from \citet{pers+79} for the
southern galaxies, and $(g\!-\!z)_{\rm SDSS}$ for NGC\,5982.  We then
determine $(V\!-\!R)_J$, $V\!-\!z_{\rm AB}$, and $(g\!-\!z)_{\rm SDSS}$ from a large
set of spectral energy distributions from the BC03 SSP models\footnote{This
  grid of models constituted  ages $\geq 5$ Gyr and $0.2 \leq Z/Z_{\odot}
  \leq 2.5$.} 
using the {\sc synphot} package within
\href{http://bit.ly/2i6m11Q}{IRAF/STSDAS}. Finally, we derive 
second-order polynomial relations between those colors to derive
$V\!-\!z_{\rm  AB}$ from $(V\!-\!R)_J$ or $(g\!-\!z)_{\rm SDSS}$. The RMS
errors of these relations are 0.008 mag between $V\!-\!z_{\rm AB}$ and
$(V\!-\!R)_J$ and 0.005 mag between $V\!-\!z_{\rm AB}$ and $(g\!-\!z)_{\rm
  SDSS}$.   

\newlength{\myone}
\settowidth{\myone}{\scriptsize 1}
\newcommand{\myl}{\hspace*{4.0pt}}

\begin{table*}[tbh]
\scriptsize
\caption{Properties of galaxies in IUE sample.}
\label{t:IUEsample}
\begin{tabular*}{\textwidth}{@{\extracolsep{\fill}}rcccccccccccc@{}} \tableline \tableline
\multicolumn{3}{c}{~~} \\ [-2.5ex]  
NGC & $({\it FUV}\!-\!V)_0$ & Mg$_2$ & [$Z$/H] & [$\alpha$/Fe] & Age & Ref. & $\sigma$ & $V_{\rm max}$ & $S_N$ & $S_{N,\,z}$ & $f_{\rm red}$ & Ref. \\
(1) & (2) & (3) & (4) & (5) & (6) & (7) & (8) & (9) & (10) & (11) & (12) & (13) \\ [0.5ex] \tableline
\multicolumn{3}{c}{~~} \\ [-2.5ex]
   221 & 4.50\plm0.17  & 0.198\plm0.002 & 0.152\plm0.030 & \llap{$-$}0.025\plm0.013  & {\myl}2.4\plm0.2  & 1 & {\myl}65\plm2 & {\myl}43\plm12 &  --- &  --- &  --- & --- \\
   224 & 3.46\plm0.17  & 0.324\plm0.002 & 0.441\plm0.048 &  0.219\plm0.017  & {\myl}7.0\plm0.8  & 1 & 154\plm4 & {\myl}74\plm40 &  --- &  --- &  --- & --- \\
   584 & 3.93\plm0.17  & 0.298\plm0.004 & 0.478\plm0.046 &  0.223\plm0.014  & {\myl}2.8\plm0.3  & 1 & 199\plm4 & {\myl}72\plm~5 &  --- &  --- &  --- & --- \\
  1399 & 2.04\plm0.17  & 0.357\plm0.008 &  0.56\plm0.12  &   0.35\plm0.04   & 10.0\plm1.4  & 2 & 332\plm5 & {\myl}37\plm22 & \llap{1}1.5\plm1.0  & 4.00\plm0.35 & 0.63 & 1 \\ 
  1404 & 3.26\plm0.17  & 0.344\plm0.007 &  0.43\plm0.09  &   0.18\plm0.03   & {\myl}8.9\plm1.3  & 2 & 227\plm4 & {\myl}96\plm19 &  2.0\plm0.5  & 0.77\plm0.20 & 0.56 & 1 \\
  1407 & 2.40\plm0.17  & 0.341\plm0.010 &  0.29\plm0.13  &   0.38\plm0.08   & 17.0\plm3.4  & 3 & 266\plm5 & {\myl}35\plm26 &  4.0\plm1.3  & 1.79\plm0.58 & 0.62 & 2 \\
  2784 & 3.65\plm0.25  & 0.334\plm0.007 &  0.67\plm0.07  &   0.23\plm0.07   & {\myl}4.5\plm0.8  & 4 & 222\plm6 & 173\plm19 &  --- &  --- &  --- & --- \\
  3115 & 3.40\plm0.17  & 0.309\plm0.006 &  0.20\plm0.06  &   0.08\plm0.05   & 17.0\plm4.7  & 4 & 259\plm3 & 106\plm{\myl}5  &  2.3\plm0.5  & 0.98\plm0.21 & 0.35 & 3 \\
  3379 & 3.82\plm0.17  & 0.329\plm0.006 & 0.299\plm0.036 &  0.259\plm0.012  & 10.0\plm1.1  & 1 & 203\plm2 & {\myl}52\plm11 &  --- &  --- &  --- & --- \\
  4472 & 3.39\plm0.17  & 0.331\plm0.005 & 0.342\plm0.046 &  0.258\plm0.021  & {\myl}9.6\plm1.4  & 1 & 282\plm3 & {\myl}50\plm20 & 5.40\plm0.57 & 2.20\plm0.23 & 0.29 & 4 \\
  4494 & 3.77\plm0.17  & 0.293\plm0.009 &  0.12\plm0.03  &   0.14\plm 0.02  & 14.0\plm2.9  & 4 & 148\plm3 & {\myl}69\plm14 &  --- &  --- &  --- & --- \\
  4552 & 2.32\plm0.17  & 0.346\plm0.006 & 0.356\plm0.034 &  0.277\plm0.011  & 12.4\plm1.5  & 1 & 250\plm3 & {\myl}~6\plm10 & 2.82\plm0.57 & 1.15\plm0.23 & 0.53 & 4 \\
  4621 & 3.14\plm0.17  & 0.355\plm0.009 &  0.65\plm0.09  &   0.30\plm 0.08  & {\myl}5.6\plm1.2  & 4 & 228\plm4 & 109\plm22 & 2.70\plm1.19 & 1.07\plm0.47 & 0.49 & 4 \\
  4649 & 2.20\plm0.17  & 0.360\plm0.006 & 0.362\plm0.029 &  0.296\plm0.012  & 14.1\plm1.5  & 1 & 331\plm5 & {\myl}55\plm22 & 5.16\plm1.20 & 2.03\plm0.47 & 0.57 & 4 \\ 
  4697 & 3.41\plm0.17  & 0.320\plm0.006 & 0.148\plm0.043 &  0.155\plm0.018  & {\myl}8.3\plm1.4  & 1 & 165\plm2 & 105\plm29 &  --- &  --- &  --- & --- \\
  4762 & 3.68\plm0.17  & 0.280\plm0.006 &  0.23\plm0.05  &   0.12\plm 0.05  & {\myl}8.8\plm2.5  & 4 & 141\plm4 & 110\plm19 &  --- &  --- &  --- & --- \\
  4889 & 2.71\plm0.17  & 0.356\plm0.008 &  0.29\plm0.03  &   0.28\plm 0.01  & 15.5\plm2.9  & 5 & 393\plm5 & {\myl}{\myl}7\plm17 &  5.5\plm0.1  & 1.97\plm0.10 & 0.60 & 5 \\ [0.5ex] \tableline 
\multicolumn{3}{c}{~~} \\ [-3ex]                                                                                                                      
\end{tabular*}
\tablecomments{Column (1): NGC number of galaxy.
  Column (2): $({\it FUV}\!-\!V)_0$ in the STMAG system from \citet{dorm+95}. 
  Column (3): Mg$_2$ index in mag from \citet{burs+88}. 
  Column (4): [$Z$/H] in dex. 
  Column (5): [$\alpha$/Fe] in dex. 
  Column (6): age in Gyr. 
  Column (7): reference of Lick index data used for stellar population parameters 
  (1 = \citealt{thom+05}, 2 = \citealt{kunt00}, 3 = \citealt{spol+06}, 4 = \citealt{trag+98},
   5 = \citealt{sanc+06}). For references other than \#1, 
   stellar population parameters were derived as described in the text. 
  Column (8): central velocity dispersion in \kms\ from LEDA. 
  Column (9): maximum rotational velocity in \kms\ from LEDA. 
  Column (10): specific frequency of GCs. 
  Column (11): specific frequency per unit $z$-band luminosity (see Equation~\ref{eq:Sn_red}). 
  Column (12): fraction of red GCs. 
  Column (13): reference of GC data (1 = this paper, 2 = \citealt{forb+06}, 3 = \citealt{jenn+14},
  4 = \citealt{peng+06}, 5 = \citealt{harr+17}). 
  } 
\end{table*}

\begin{table*}[tbh]
  \begin{center}
\scriptsize
\caption{Relevant properties of galaxies in SAURON sample with $S_N$ data.}
\label{t:SAURONsample}
\begin{tabular*}{12cm}{@{\extracolsep{\fill}}rcrcccccc@{}} \tableline \tableline
\multicolumn{3}{c}{~~} \\ [-2.5ex]  
NGC & $({\it FUV}\!-\!V)_{\rm AB,\,0}$ & \multicolumn{1}{c}{Age} & $\sigma$ & $V_{\rm max}$ & $S_N$ & $S_{N,\,z}$ & $f_{\rm red}$ & Ref. \\
(1) & (2) & \multicolumn{1}{c}{(3)} & (4) & (5) & (6) & (7) & (8) & (9) \\ [0.5ex] \tableline
\multicolumn{3}{c}{~~} \\ [-2.2ex]
    474 &   6.76\plm0.19 &  8.89\plm1.54 & 154\plm3  & {\myl}30\plm{\myl}6 & {\myl}2.10\plm0.50 &  0.99\plm0.24 &  0.13 & 1 \\
   4387 &   6.98\plm0.14 & 10.34\plm1.79 & 100\plm3  & {\myl}58\plm{\myl}6 & {\myl}1.52\plm0.21 &  0.65\plm0.09 &  0.23 & 2 \\
   4473 &   6.90\plm0.06 & 11.76\plm2.03 & 179\plm3  & {\myl}57\plm{\myl}6 & {\myl}1.98\plm0.51 &  0.88\plm0.23 &  0.43 & 2 \\
   4486 &   5.36\plm0.07 & 17.70\plm2.04 & 323\plm4  & {\myl}20\plm21 & 12.59\plm0.77 &  4.19\plm0.25 &  0.27 & 2 \\ 
   4570 &   6.67\plm0.05 & 12.82\plm2.22 & 187\plm5  &   ---     & {\myl}1.09\plm0.18 &  0.53\plm0.09 &  0.36 & 2 \\
   4621 &   6.32\plm0.06 & 13.67\plm2.36 & 228\plm4  & 109\plm22 & {\myl}2.70\plm1.19 &  1.07\plm0.47 &  0.49 & 2 \\
   4660 &   6.80\plm0.23 & 12.55\plm2.05 & 192\plm3  & 145\plm14 & {\myl}2.97\plm0.41 &  1.12\plm0.15 &  0.13 & 2 \\
   5982 &   6.24\plm0.07 &  8.90\plm0.90 & 242\plm4  & {\myl}46\plm26 & {\myl}2.60\plm0.60 &  1.06\plm0.25 &  0.48 & 1 \\ [0.5ex] \tableline
\multicolumn{3}{c}{~~} \\ [-3ex]                                                                                                                      
\end{tabular*}
\tablecomments{Column (1): NGC number of galaxy.
  Column (2): $({\it FUV}\!-\!V)_{0}$ in AB mag from \citet{bure+11}.
  Column (3:) age in Gyr from \citet{kunt+10}. 
  Column (4): central velocity dispersion in \kms\ from LEDA. 
  Column (5): maximum rotation velocity in \kms\ from LEDA. 
  Column (6): specific frequency of GCs. 
  Column (7): specific frequency per unit $z$-band luminosity (see Equation~\ref{eq:Sn_red}). 
  Column (8): fraction of red GCs. 
  Column (9): reference of GC data (1 = \citealt{sikk+06}, 2 = \citealt{peng+08}).
}
\end{center}
\end{table*}

\begin{figure*}
\centerline{\includegraphics[width=15.5cm]{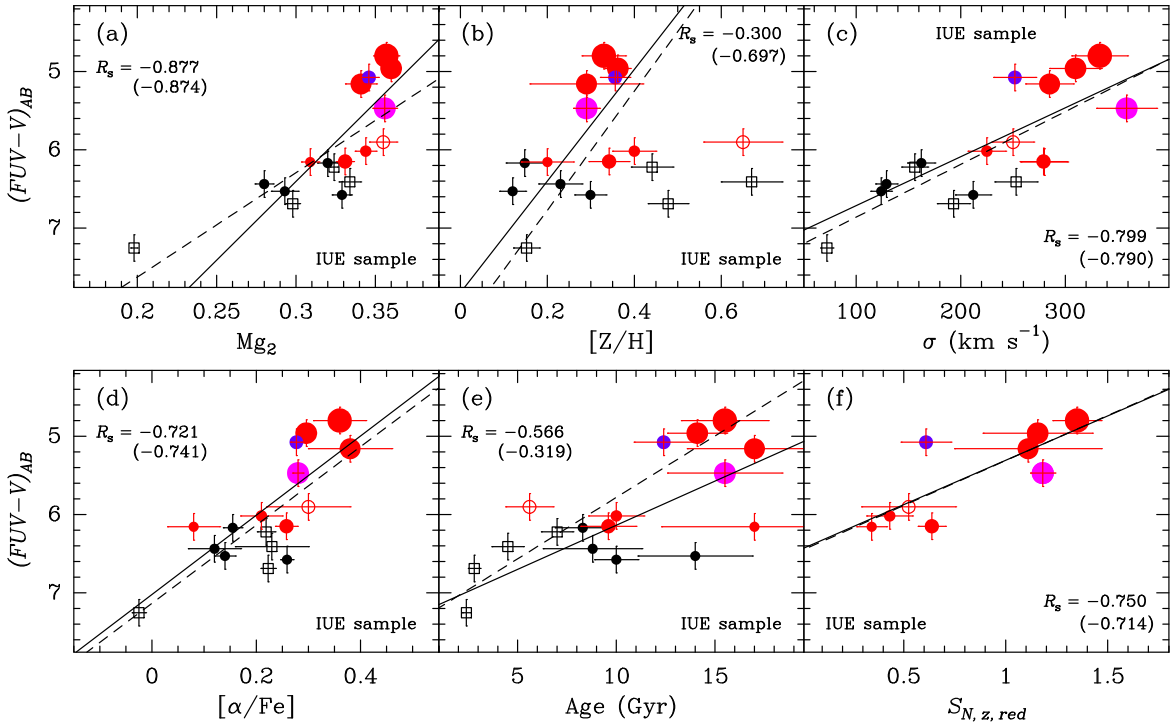}}
\caption{Correlations between \FUVmV\ and galaxy properties for the
  IUE sample. \emph{Panel (a)}: \FUVmV\ vs.\ Mg$_2$. \emph{Panel (b)}:
  \FUVmV\ vs.\ [$Z$/H]. \emph{Panel (c)}: \FUVmV\ vs.\ central
  velocity dispersion. \emph{Panel (d)}:
  \FUVmV\ vs.\ \aFe. \emph{Panel (e)}: \FUVmV\ vs.\ age. \emph{Panel
    (f)}: \FUVmV\ vs.\ $S_{N,\,z,\,{\rm red}}$, the specific frequency
  of metal-rich GCs. In each panel, red, purple, or magenta symbols
  refer to galaxies for which $S_{N,\,z,\,{\rm red}}$ is available; the size of
  those symbols scales linearly with the value of $S_{N,\,z,\,{\rm red}}$. 
  The purple symbol indicates NGC 4552, a galaxy with a LINER nucleus
  emitting part of its FUV luminosity. The magenta symbol indicates NGC 4889, a
  Coma elliptical whose distance is $\ga 4$ times larger than that of all other 
  galaxies in this sample. 
  Open symbols refer to galaxies with SSP-equivalent ages
  $\leq 8$ Gyr, while filled symbols refer to older galaxies. The Spearman
  rank correlation coefficient $R_s$ is mentioned in each panel for the
  relation in question, both for all galaxies and for galaxies with
  SSP-equivalent ages $>$~8 Gyr (the latter in parentheses). Linear
  least-square fits to the data are shown in each panel by dashed lines (for
  all galaxies) and solid lines (for galaxies with SSP ages $>$~8 Gyr). See
  discussion in Sect.\ \ref{s:rela}. } 
\label{f:FUVmVfig1}
\end{figure*}

Figure~\ref{f:FUVmVfig1} shows \FUVmV\ versus the
various stellar population parameters, the central velocity dispersion from
\href{http://leda.univ-lyon1.fr/}{HyperLeda}, and $S_{N,\,z,\,{\rm red}}$ for the
IUE sample. Note that the $m(1550)\!-\!V$ magnitudes from \citet{dorm+95} were
transformed to \FUVmV\ in the AB magnitude system using 
$({\it FUV}\!-\!V)_{\rm AB} = m(1550)\!-\!V + 2.75$. 
The Spearman rank correlation coefficient $R_s$ is
mentioned for each relation in the various panels.  
$R_s$ is measured and shown twice for each relation: once for all galaxies in
the sample, and once for the galaxies with SSP-equivalent ages of $\geq 8$
Gyr. This was done because the latter is thought to be the approximate age at 
which single EHB stars start producing the UV upturn feature as observed 
\citep[e.g.,][]{chun+17}, and EHB stars are generally recognized to 
dominate the FUV luminosity of giant ETGs.   

\begin{figure}
  \vspace*{1mm}
\centerline{\includegraphics[width=6.cm]{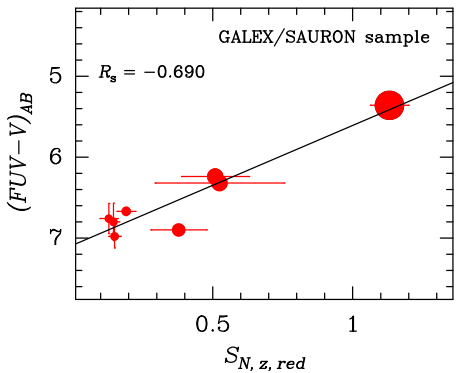}}
\caption{Same as panel (f) of Figure~\ref{f:FUVmVfig1}, but now for the SAURON sample.}
\label{f:FUVmVfig2}
\end{figure}


Focusing on panels (a), (b), (d), and (e) of Figure~\ref{f:FUVmVfig1}
(i.e., the stellar population parameters for the IUE sample), it can be seen
that the well-known anticorrelation between \FUVmV\ and Mg$_2$ does not
translate to a significant anticorrelation between \FUVmV\ and \ZH\ when taking
all galaxies into account. However, the main outliers in this relation all have
SSP-equivalent ages $< 8$ Gyr and high \ZH, and \FUVmV\ does anticorrelate
with \ZH\ when considering only the galaxies with $\geq 8$ Gyr. 
Panel (e) reveals a significant anticorrelation between
\FUVmV\ and SSP-equivalent age when considering all galaxies; however, this
relation seems to be largely driven by the galaxies with ages $< 8$ Gyr,
since the correlation weakens quite significantly when only considering the
older galaxies.
In this context, we remind the reader that the determinations of SSP-equivalent
age of galaxies employed here are based on luminosity-weighted measurements of
line indices in the blue and visual parts of the spectrum. Such age measurements
are quite sensitive to the presence of  young populations with small mass
fractions \citep[e.g.,][]{thom+03}. As such,  SSP-equivalent ages of galaxies
formally measure lower limits to the average age. 
Finally, the anticorrelation between \FUVmV\ and
\aFe\ shown in panel (d) is stronger than those between \FUVmV\ and either
\ZH\ or SSP age. This suggests that, among the stellar population parameters,
\aFe\ seems to be the strongest driver of the \FUVmV\ vs.\ Mg$_2$
anticorrelation, even though it seems probable that \ZH\ and age also have
some influence. 
Panel (c) of Figure~\ref{f:FUVmVfig1} shows the well-known strong
anticorrelation of \FUVmV\ with central velocity dispersion, and hence with the
depth of the potential well of the galaxy (and the surrounding galaxy group or
cluster if present). 

The trends and anticorrelations mentioned above are generally consistent with
those found by other recent studies of the UV upturn among ETGs 
\citep[e.g.,][]{bure+11,cart+11,jeon+12,smit+12a}, except that the
anticorrelations found in the current paper are typically stronger than those in
the other studies. 
We suggest that this may be due to the FUV emission in ETGs being 
more centrally concentrated than the $V$-band light \citep{ohl+98,cart+11}. Thus, 
smaller measurement apertures in conjunction with aperture matching between 
measurements of the various population properties are likely to reveal 
stronger correlations between \FUVmV\ and other parameters. The IUE 
measurements were made through an aperture of $10'' \times 20''$ 
which corresponds to a (circularized) radius in the approximate range 
$R_{\rm eff}/8$ to $R_{\rm eff}/4$ for most galaxies in the IUE
sample\footnote{NGC 4889 is an exception due to its
  distance being $\ga 4$ times larger than that of all other galaxies in the IUE 
  sample. Hence it is assigned a different symbol in Figure~\ref{f:FUVmVfig1}.} 
\citep[see, e.g.,][]{caon+93,goud+94a}. In contrast, the studies of the UV 
upturn in the SAURON sample \citep{bure+11,jeon+12} used measurements of 
\FUVmV\ within an aperture of $R_{\rm eff}/2$ for their correlation analysis,
so that the influence of the FUV-emitting population can be expected to be
somewhat diluted relative to the \FUVmV\ measurements of galaxies in the IUE
sample.  

A new relation found here is that \FUVmV\ anticorrelates strongly with
$S_{N,\,z,\,{\rm red}}$, the specific frequency of metal-rich GCs.  This is
shown in panel (f) of Figure~\ref{f:FUVmVfig1} for the IUE sample, and in 
Figure~\ref{f:FUVmVfig2} for the SAURON sample. The nature and possible
implications of this relation are discussed in the next section. 

\section{GCs as a Source of far-UV flux in Galactic Spheroids} 
\label{s:disc} 

The anticorrelation of \FUVmV\ with $S_{N,\,z,\,{\rm red}}$ reported here
could in principle simply reflect the already known anticorrelation 
between \FUVmV\ and $\sigma$ in conjunction with the correlation between $M_V$
and $S_{N,\,z,\,{\rm red}}$ reported by \citet{peng+08}
for galaxies in the Virgo cluster with $M_z \la -21$. 
However, the correlation between \FUVmV\ and $S_{N,\,z,\,{\rm red}}$ is stronger than
that between $M_V$ and $S_{N,\,z,\,{\rm red}}$. Furthermore, panels
(a)\,--\,(e) of Figure~\ref{f:FUVmVfig1} show that galaxies with the highest
values of $S_{N,\,z,\,{\rm red}}$ systematically have bluer \FUVmV\ colors than that
indicated by the linear fits to the relations shown by the lines in each
panel. At face value, this seems to suggest that $S_{N,\,z,\,{\rm red}}$ has a
\emph{causal} anticorrelation with \FUVmV.
With this in mind, 
we consider the hypothesis that the He-enhanced
populations that are likely responsible for the UV upturn are produced
in massive star clusters and subsequently disperse slowly into the
field population of galaxies by means of \emph{dissolution of
  metal-rich GCs in the strong tidal field within the inner regions of
  luminous ETGs}. This hypothesis is explored below. 

\subsection{Long-term Dissolution of Metal-Rich GCs}
\label{sub:GCdiss}

\subsubsection{Empirical Evidence}

First, we describe empirical evidence for significant amounts of mass loss from
metal-rich GCs in the central regions of massive ETGs. Several
studies of GC systems around ETGs have shown that surface
number density profiles of metal-rich GCs outside $\sim$\,1 $R_{\rm eff}$
typically follow the surface brightness profile of the host galaxy, while the
metal-poor GCs have a much shallower profile
\citep[e.g.,][]{geis+96,forb+98,rhozep04,dirs+05,bass+06,harr+10,stra+11}.
As such, metal-rich GCs are thought to be physically associated with the
luminous component of the galaxies (``bulge''), while the metal-poor GCs are
more of a ``halo''-like population. Recent studies of nearby ETGs 
have shown that this difference in radial distribution between
metal-rich and metal-poor populations is shared by the field
stars \citep{harr+07,rejk+14,peac+15}. In fact, these studies found the
fraction of metal-poor field stars in the inner regions of ETGs 
to be even lower than one would predict based on the difference in radial
number density profiles between metal-rich and metal-poor GCs mentioned above. 

However, \emph{HST} studies of ``normal,'' old ETGs have shown
that the surface number densities of metal-rich GCs are significantly 
depleted \emph{in the central regions} relative to the galaxy light profile
\citep[e.g.,][]{forb+06,peng+06,harr+17}. Interestingly, this depletion is
much weaker for early-type merger remnant galaxies such as NGC~1316
\citep{goud+01} and NGC~3610 \citep{goud+07}, which are only a few Gyr
old, so that dynamical evolution of GCs has had much less time to disperse
their stars into the field. It thus seems fair to suggest that these
depletions of the surface number densities of metal-rich GCs toward the galaxy
centers are due to dynamical evolution (two-body relaxation (or ``evaporation'') 
and tidal shocking) in the central regions.

\subsubsection{Quantitative Estimates}

To estimate the amount of GC mass loss occurring in the inner regions of
ETGs, we proceed as follows. We assume that GCs are tidally
limited, meaning that their 
average mass densities are determined by the galaxy density inside their
orbits \citep[e.g.,][]{king62}. For simplicity, we assume circular orbits, and
we follow \citet{fz01} by relating the mass densities of GCs to their
galactocentric radius $R_{\rm gal}$ in a steady-state isothermal potential
$\Phi(R_{\rm gal}) = V_{\rm gal}^2 \ln\, (R_{\rm gal})$ with a fixed maximum
velocity $V_{\rm gal} = (V^2 + \sigma^2)^{1/2}$ where $V$ is the galaxy's
maximum rotational velocity and $\sigma$ is its central velocity dispersion. In
this case, the mean evaporative mass loss rate $\mu_{\rm ev}$ of a cluster is
given by
\begin{equation}
  \mu_{\rm ev} \; \simeq \; 2.9 \times 10^4 \; \left(\frac{R_{\rm
      gal}}{\rm kpc}\right)^{-1}  \left(\frac{V_{\rm gal}}{220\; {\rm km\,s}^{-1}}\right)
  \; M_{\odot}\;{\rm Gyr}^{-1}
\label{eq:mu}
\end{equation}
\citep[see Equations (4) and (15) in][]{fz01}. It should be emphasized that
$\mu_{\rm ev}$ in Equation\ (\ref{eq:mu}) likely represents a lower limit to
the actual average mass loss rate of surviving metal-rich GCs since their
birth. Under the assumption that metal-rich GCs were formed in situ during the
star-forming era of massive building blocks of present-time giant ETGs, it is
likely that disruption rates of GCs in such dense environments 
at high redshift were significantly higher than they are currently, due to the
stronger tidal perturbations \citep[this may especially be the case for the
  metal-rich GCs, see][]{krui15}. Furthermore, application
of Equation~(\ref{eq:mu}) neglects early mass loss by processes such as
residual gas expulsion \citep{baum+08} and ejection of stars from the cluster
as it expands adiabatically after the death of the massive stars in the
cluster \citep{hill80}.
The rate of the latter type of mass loss is strongly dependent on the level of 
primordial mass segregation of the cluster \citep[e.g.,][]{vesp+09}.

\subsection{Comparison with Radial Number Density Distributions}
\label{sub:radprof}

\begin{figure*}
\centerline{\includegraphics[width=13.5cm]{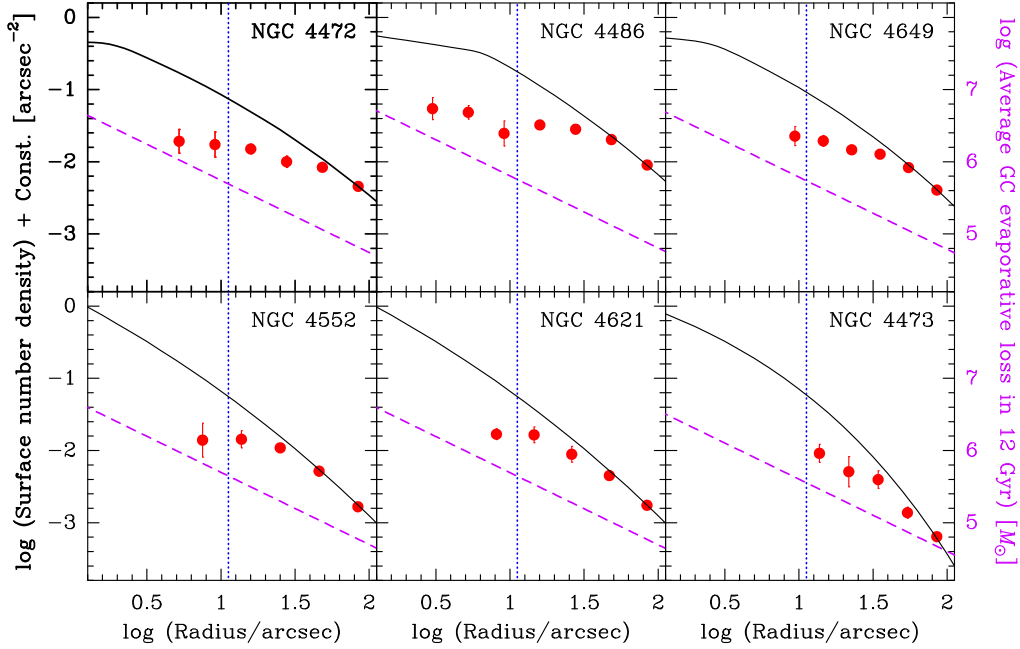}}
\caption{Radial surface number density profiles for the metal-rich GCs in six 
  giant ETGs in the Virgo cluster of galaxies from
  \emph{HST}/ACS data. The galaxy names are indicated at the top right of each
  panel. Surface number density data were derived as explained in the text, and
  are shown as open circles. 
  The solid lines represent the surface brightness profiles of the 
  integrated $z$-band light of the galaxies (taken from \citealt{ferr+06}), 
  normalized to agree with the value of the outermost GC number density. The purple
  dashed lines represent the logarithm of the average mass accumulated by
  evaporation of GCs over a time span of 12 Gyr as a function of galactocentric
  radius according to Equation~(\ref{eq:mu}), for which the scale is shown on
  the right-hand side of the figure. For comparison purposes, the circularized 
  radius of the IUE measurement aperture is indicated by vertical blue dotted
  lines. See discussion in Section~\ref{sub:radprof}.  
  }
\label{f:radprof}
\end{figure*}

We compare the mass lost from GCs as a function of galactocentric radius as
described above with the radial number density distributions of metal-rich GCs
for six giant ETGs from the ACS Virgo Cluster Survey
\citep[ACSVCS;][]{cote+04,jord+09}.  
These galaxies were selected for hosting GCs whose $g-z$ color distribution
clearly shows two distinct peaks, facilitating a simple distinction between
metal-poor and metal-rich GCs (see \citealt{goukru13} for the exact selection
criteria). Four of these galaxies are in the IUE sample
described in Section~\ref{s:rela}, while the other two (NGC~4473 and NGC~4486 =
M87) are in the SAURON sample. 
Radial distributions of surface number densities of metal-rich GCs in these
galaxies were derived from the tables in \citet{jord+09} as follows. GCs were
first defined as sources to which \citet{jord+09} assigned a probability of
being a GC of $p_{\rm GC} \geq 0.5$. To avoid incompleteness-related issues, we
considered only GCs with $z \leq 22.9$, corresponding to the 90\% completeness
limit in the inner regions of the host galaxy with the highest central surface
brightness. Completeness-corrected numbers of metal-rich GCs, $N_{\rm GC}$, were
determined in annuli with logarithmic spacing in galactocentric radius. These
$N_{\rm GC}$ values were then divided by the area of the annuli, taking into
account the limited azimuthal coverage of the outermost annuli.
Figure~\ref{f:radprof} shows the surface number density distributions of the
metal-rich GCs in these galaxies, along with the surface brightness
profiles of the parent galaxy and the average amount of mass lost from GCs
as a function of galactocentric radius during a time span of 12 Gyr according to
Equation~(\ref{eq:mu}) for comparison purposes. Note that the surface number
density profiles of metal-rich GCs flatten out significantly (relative to the
surface brightness profile of the parent galaxy) inside
$R_{\rm gal} \approx 30''$ of these galaxies, corresponding to $\approx 2.5$
kpc at the distance of the Virgo cluster (16.5 Mpc, see \citealt{jord+09}),
where the typical evaporative mass lost from GCs over 12 Gyr according to
Equation~(\ref{eq:mu}) (hereafter referred to as $\Delta_{\rm GC}$) is of
order $2\times10^5\;M_{\odot}$.
Furthermore, the level of depletion of the surface number densities of
metal-rich GCs relative to the galaxy light increases toward smaller
$R_{\rm gal}$. This is consistent with the fact that $\Delta_{\rm GC}$
increases to $10^6\; M_{\odot}$ or beyond in those innermost regions: only the
GCs with the highest initial masses have been able to survive dissolution by
the strong tidal forces in those regions until the present time.

\section{Feasibility of Proposed Scenario} 
\label{s:feas} 

\subsection{Energetics}
\label{sub:ener}

In this section we evaluate the feasibility of the hypothesis that He-rich EHB
stars associated with (now mostly dissolved) massive metal-rich GCs can indeed 
produce the observed range of \FUVmV\ among giant ETGs. To do so,
we use the observational results of the FUV emission of massive GCs in M87
from \citet{sohn+06}. 

\subsubsection{Required Numbers of GCs}
  
\begin{figure}
\centerline{\includegraphics[width=6.5cm]{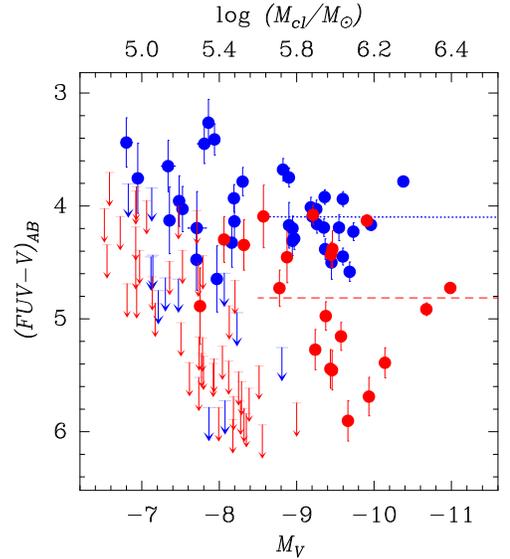}}
\caption{(\FUVmV)$_{\it AB}$ versus $M_V$ for GCs in M87 from
  \citet{sohn+06}. Blue symbols indicate metal-poor GCs and red symbols
  indicate metal-rich GCs. FUV upper limits are indicated by downwards
  arrows. The blue dotted line and red dashed line represent linear
  least-square fits to the metal-poor and metal-rich GCs with $M_V \leq -8.5$, 
  respectively. Estimated GC masses are indicated along the top abscissa.
  }
\label{f:FUVmV_vs_mass}
\end{figure}

We first make the assumption that the FUV properties of the surviving
metal-rich GCs constitute a suitable proxy for the GCs that dissolved over
the last several Gyr. Since it is likely that the initial masses of the
dissolved GCs were on average lower than those of the surviving GCs (at a
given tidal field strength and evaporation rate, that is), the implicit
assumption is that the dependence of \FUVmV\ on GC mass is small within the
applicable range of GC masses. To assess the validity of this assumption, we
plot \FUVmV\ versus $M_V$ for the M87 GCs from \citet{sohn+06} in
Figure~\ref{f:FUVmV_vs_mass}.\footnote{Following \citet{sohn+06}, we use
  $({\it FUV}\!-\!V)_{\rm AB} = ({\it FUV}\!-\!V)_{\rm STMAG} + 2.82$.} To
discriminate between metal-poor and metal-rich GCs in this context, we adopt
the color threshold $V\!-\!I$ = 1.09 as determined by \citet{kund+99} whose
\emph{HST}/WFPC2 photometry was used by \citet{sohn+06} as well. Current GC
mass estimates are indicated in Figure~\ref{f:FUVmV_vs_mass}, assuming a
typical ${\cal{M}}/L_V = 1.8$ \citep{goufal16,harr+17}. Note that
there is no significant mass dependence of \FUVmV\ for GCs with
$M_V \la -8.0$, corresponding to ${\cal{M}} \ga 2 \times 10^5\;M_{\odot}$. 
This is the case for both the metal-poor and the metal-rich GCs. As such,
the following calculations pertain to GCs with current masses
${\cal{M}} \ga 2 \times 10^5\;M_{\odot}$.  
Among these massive metal-rich GCs in M87, we find a mean $M_V^0 = -9.64$,
corresponding to a mean mass ${\cal{M}} \sim 1 \times 10^6\; M_{\odot}$. Their 
mean \FUVmV\ = 4.82 in AB mag, corresponding to $M_{\it FUV}^0 = -4.82$ in AB mag
at the distance of the Virgo cluster. 

A second assumption we make is that single He-rich EHB stars created in massive
GCs are unlikely to be the \emph{only} source of FUV luminosity in ETGs. For 
instance, the model of \citet{han+07}, i.e., hot helium-burning stars that have
lost their hydrogen-rich envelopes in binary interactions (be it in GCs or
in the field), may well provide a \emph{minimum} strength of the UV
upturn (or maximum \FUVmV, hereafter $({\it FUV}\!-\!V)_{\rm max}$) in
ETGs. The model of \citeauthor{han+07}\ predicts $({\it FUV}\!-\!V)_{\rm AB}$ in
the range $6.5 - 7.0$ (with the precise value depending on the parameter values 
of the model), with a negligible dependence on age for ages $\ga 1$ Gyr.
Interestingly, this range of $({\it FUV}\!-\!V)_{\rm max}$ is roughly
consistent with the maximum  value of \FUVmV\ seen in samples of ETGs 
\citep[see Figure~\ref{f:FUVmVfig1};][]{bure+11,jeon+12}. 
With this in mind, we explore to what extent the \emph{range} in
\FUVmV\ seen among ETGs can be understood in terms of GC formation efficiencies
and dynamical evolution of massive GCs.  
  
The range of UV upturn strengths among ETGs in \FUVmV\ is $\sim$\,1.5 mag 
\citep[see Figure~\ref{f:FUVmVfig1};][]{bure+11}. Relevant ETGs
representing the bottom and top of this range are NGC 4473 and NGC 4649,
respectively.\footnote{We exclude the strong UV upturns in NGC 1399 and M87 from 
  this exercise, since their FUV emission is due in part to unresolved
  nuclear emission \citep{cart+11}.}  These two galaxies have the same SSP age
to within the uncertainties as determined from their Lick indices
(see Tables~\ref{t:IUEsample} and \ref{t:SAURONsample}), 
suggesting that their production rates of EHB stars by stellar evolution
are consistent with each other as well. 
Since NGC 4473 has \FUVmV\ = $6.90 \pm 0.06$ within 
$R_{\rm eff}/2$ \citep{bure+11}, it constitutes a relevant example of an ETG
for which a significant fraction of the FUV luminosity may be due to hot
subdwarf stars produced in binary systems. 
To derive \FUVmV\ within $R_{\rm eff}/2$ for NGC 4649, we download GALEX
dataset GI3\_041008\_NGC4621 from the
\href{http://galex.stsci.edu/GR6/}{STScI/MAST GR6 archive}, with an exposure
time of 1658 s in the NUV and FUV passbands.
This GALEX dataset is available at the MAST archive at
\dataset[10.17909/T9V68G]{doi.org/10.17909/T9V68G}. Following \citet{bure+11}, we
carry out surface photometry of NGC 4649 using the {\sc ellipse} task in
\href{http://bit.ly/2i6m11Q}{IRAF/STSDAS}, using the NUV image to perform
ellipse fitting followed by imposing those ellipses on the FUV
image. For the sky background level we use the mean of unclipped mean
values in 15 square apertures located around the galaxy, in regions free from
stars or other galaxies. We adopt a photometric zero point of 18.82 mag
\citep{morr+05}, and correct the photometry for Galactic extinction using the
\ebv\ value from \citet{schl+98} as mentioned in
\href{http://ned.ipac.caltech.edu}{NED} in conjunction with the extinction law
of \citet{card+89}. 
To obtain $V$-band photometry for NGC 4649, we use the $B$-band surface
photometry tables of 
\citet{pele+90} in conjunction with the $B\!-\!V$ colors from \citet{burs+87},
interpolating in galactocentric radius. Adopting $R_{\rm eff}$ = 74\farcs3 for
NGC 4649 \citep{burs+87}, we obtain $M_{\it FUV,\,AB}^0 = -15.50$,
$M_V^0 = -21.01$, and \FUVmV\ = 5.51 within $R_{\rm eff}/2$, i.e., 1.39 mag
bluer in \FUVmV\ than NGC 4473.

To determine how many ``average'' FUV-bright massive metal-rich GCs as
seen in M87 are needed to produce the ``extra'' FUV luminosity 
in NGC 4649 relative to NGC 4473, we iteratively add such GCs (for which
$M_V^0 = -9.64$ and $M_{\it FUV}^0 = -4.82$, see above) to NGC 4473 and
re-evaluate its integrated $M_V$ and $M_{\it FUV}$ until we produce a galaxy
with \FUVmV\ = 5.51. We find that this would require 7789 such GCs. These would 
produce $M_V = -19.41$, which is 1.60 mag fainter than the actual light within 
$R_{\rm eff}/2$ in NGC 4649, equivalent to a $V$-band luminosity fraction
$f_V = 0.23$.

In conclusion, the range in \FUVmV\ seen among ETGs can be produced by 
(currently largely dissolved) FUV-bright GCs similar to those seen in M87 with
numbers of such GCs that produce $V$-band luminosity fractions up to 
$f_V \sim 0.23$.

\subsubsection{Consistency with GC Mass Loss Scenario}

One might wonder whether this range of $f_V$ is consistent with that
expected from GC mass loss over a time span similar to the age of such
galaxies (for which we adopt 12 Gyr as before). To address this question, we
use the semi-analytical dynamical evolution model of \citet{goufal16}. As
commonly done in star cluster evolution modeling, we adopt a \citet{sche76}
function for the initial cluster mass function (ICMF): 
\begin{equation}
\psi_0 \, (\cM_0) = A \; \cM_0^{\beta} \; \exp\,(- \cM_0/\cM_{\rm c})\mbox{,}
\label{eq:schech}
\end{equation}
where $\cM_0$ is the initial cluster mass, and the adjustable parameters are
$A$, $\beta$, and $\cM_{\rm c}$. The ICMF has a power-law shape with
exponent $\beta = -2$ below $\cM_{\rm c}$ to mimic the observed mass
functions (MFs) of young cluster systems
\citep[e.g.,][]{zhafal99,chan+10,port+10,whit+14}, and it has an exponential
decline above the ``characteristic truncation mass'' $\cM_{\rm c}$ as
suggested by the observed tail of the GC mass function (GCMF) at
$\cM \ga 10^6 \; M_{\odot}$ for GC systems in ancient galaxies. 
Following \citet{goufal16}, we incorporate mass loss by stellar evolution based
on the BC03 models and we make the common assumption that the long-term
dynamical mass loss of GCs is dominated by two-body relaxation \citep[see also,
  e.g.,][]{gneost97,dine+99,krui15}. In that case, the evolving GCMF at time
$t$ is  
\begin{eqnarray}
\psi\,(\cM,\,t) \; & = & \; A\;\cM^{\gamma - 1} \; (\cM^{\gamma} + \mu_{\rm ev} \,
 t)^{(\beta - \gamma + 1)/\gamma} \nonumber \\
  & & \qquad \times \: \exp\left[- \, (\cM^{\gamma} + \mu_{\rm ev} \,
    t)^{1/\gamma}/\cM_{\rm c} \right] 
\label{eq:finMF}
\end{eqnarray}
\citep{goufal16} where $\mu_{\rm ev}$ is the evaporative mass loss rate
(cf.\ Equation \ref{eq:mu}) and $\gamma \leq 1$ denotes the exponential
dependence of the dissolution  time on GC mass
(i.e., $t_{\rm dis} \propto \cM^{\gamma}$).  
$\gamma = 1$ for classical two-body relaxation as defined by \citet{spit87}
and others. While the actual value of $\gamma$ is currently not well
constrained for GCs with $\cM \ga 10^5\; M_{\odot}$ after a Hubble time of
dynamical evolution \citep[see discussion in][]{goufal16}, the main effect of
$\gamma < 1$ in the context of this study is that it yields slightly lower
mass loss rates for low-mass clusters (i.e., GCs with $\cM < \mu_{\rm ev}\,t$)
relative to the case of $\gamma = 1.$\footnote{This effect is not detectable in
  observations because the lower mass loss rates for those low-mass GCs result
  in higher $\cM/L$ ratios, and the two effects cancel out in luminosity
  functions \citep{goufal16}.} 
To maximize the mass loss rate for low-mass GCs relative to high-mass GCs,
we choose $\gamma = 1$. As such, the values for $f_V$ calculated below 
formally represent lower limits. 

\begin{figure}
\centerline{\includegraphics[width=7.5cm]{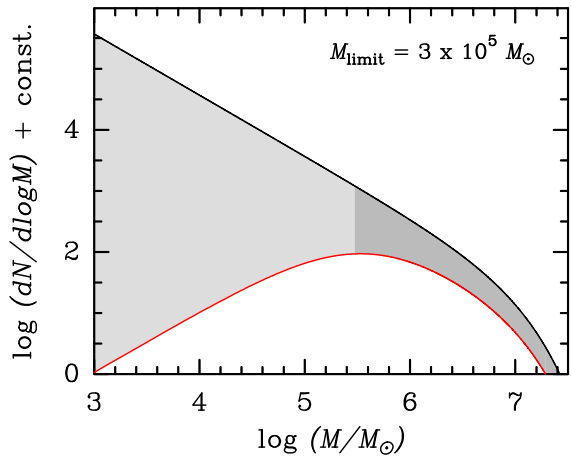}}
\caption{Example illustration of the determination of $f_{\rm GC}$, the
  fraction of mass lost from GCs with initial masses above a certain mass limit 
  $\cM_{\rm limit}$ over a time span of 12 Gyr. 
  The black solid line represents the ICMF (i.e., Equation \ref{eq:schech})
  using $\cM_{\rm c} = 10^7\; M_{\odot}$, 
  while the red solid line depicts the GCMF after $t$ = 12 Gyr of dynamical
  evolution (Equation \ref{eq:finMF} with $\gamma = 1$), 
  using an evaporation rate $\mu_{\rm ev} = 3\times10^4\; M_{\odot}\;{\rm Gyr}^{-1}$. 
  The hashed region indicates the amount of mass lost by dynamical evolution
  over 12 Gyr, while the darker gray part of that region indicates the fraction
  of mass lost from GCs with initial masses
  $\cM_0 \geq \cM_{\rm limit} = 3\times10^5\;M_{\odot}$.
  See discussion in Section~\ref{sub:ener}.2.  
  }
\label{f:MFexample}
\end{figure}

To convert $f_V$ to a mass fraction $f_{\rm GC}$ at $t$ = 12 Gyr, we assume
a GC mass-independent $\cM/L_V = 1.8 \; M_{\odot}/L_{V,\,\odot}$, which is a good
approximation to dynamical $\cM/L_V$ values determined by various studies
of stellar kinematics in GCs (see \citealt{goufal16}, \citealt{harr+17}, and
references therein). 
We then determine $f_{\rm GC}$ for GCs above a given initial mass
$\cM_{\rm limit}$ by integrating the ICMF (i.e., Equation
\ref{eq:schech}) as well as Equation (\ref{eq:finMF}) for $t$ = 12 Gyr and
evaluating the difference. This procedure is illustrated in
Figure~\ref{f:MFexample}. This evaluation is done for several values of
$\cM_{\rm limit}$ while varying $\mu_{\rm ev}$ and $\cM_{\rm c}$ within ranges
typically found among GCMF studies of massive galaxies 
\citep[$4.0 \leq \log\,(\mu_{\rm ev}/(M_{\odot}\;{\rm Gyr}^{-1})) \leq 5.5$ and
$6.0 \leq \cM_{\rm c}/M_{\odot} \leq 7.5$,
  see][]{jord+07,mclfal08,chan+10,bast+12,goufal16,john+17}.

\begin{figure*}
\centerline{\includegraphics[width=15.cm]{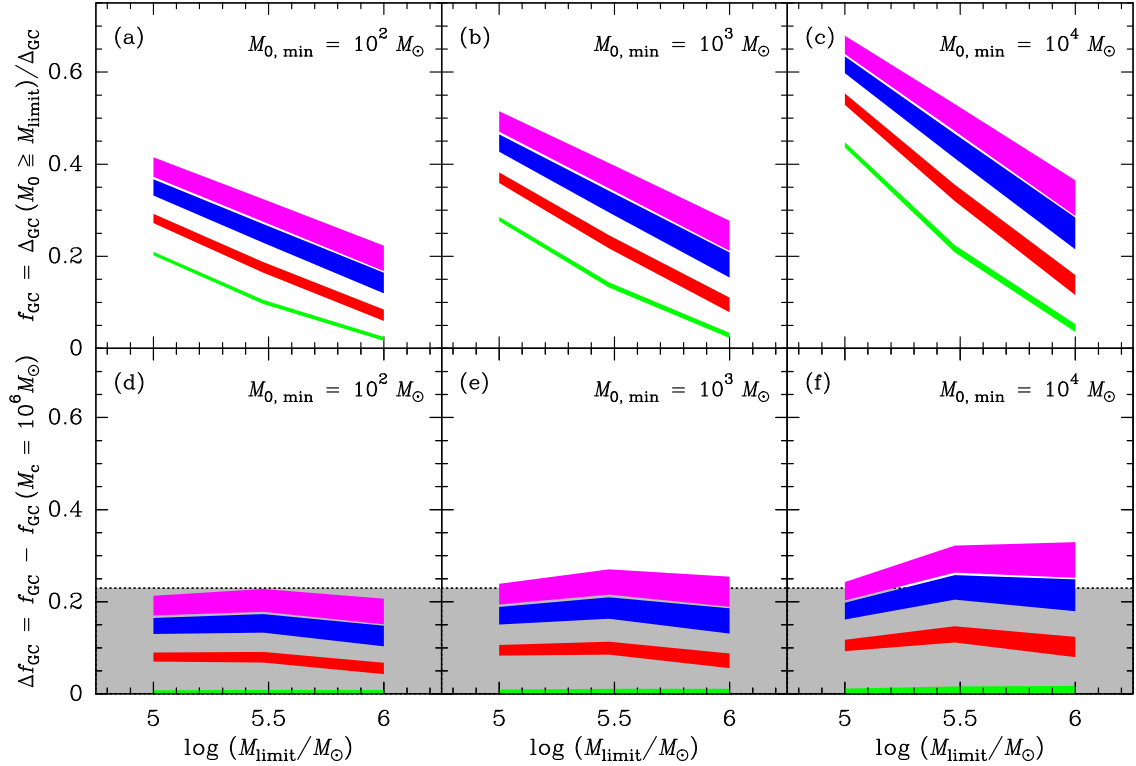}}
\caption{\emph{Top panels}: $f_{\rm GC}$ versus $\cM_{\rm limit}$. 
  The colored regions in each panel show results obtained by
  numerically integrating Equations (\ref{eq:schech}) and (\ref{eq:finMF}) for
  $4.0 \leq \log\,(\mu_{\rm ev}/(M_{\odot}\;{\rm Gyr}^{-1})) \leq 5.5$ and
  different values of $\cM_{\rm c}$. Green regions represent
  $\log\,(\cM_{\rm c}/M_{\odot}) = 6.0$, red regions represent
  $\log\,(\cM_{\rm c}/M_{\odot}) = 6.5$,   blue regions represent
  $\log\,(\cM_{\rm c}/M_{\odot}) = 7.0$, and  magenta regions represent
  $\log\,(\cM_{\rm c}/M_{\odot}) = 7.5$. 
  The different panels indicate different assumptions for the low-mass
  truncation of the ICMF, as indicated at the top right of each panel. 
  \emph{Bottom panels}: $\Delta f_{\rm GC}$ versus $\cM_{\rm limit}$. The gray region
  depicts the range of $f_{\rm GC}$ implied by the hypothesis that the range of
  \FUVmV\ seen among ETGs is caused by dissolution of massive metal-rich GCs in
  the inner regions,
  based on the FUV luminosities of M87 GCs by \citet{sohn+06}. 
  See discussion in Section~\ref{sub:ener}.2.  
  }
\label{f:Mfracplot}
\end{figure*}

For the low-mass truncation of the power-law tail of the ICMF, many studies
use $\cM_{0,\,{\rm min}} = 10^2 \; M_{\odot}$ which corresponds to the
approximate minimum mass of a cluster for which the evaporation time scale is
$10^8$ yr, the typical lifetime of open clusters in the solar
neighborhood \citep{adamye01,ladlad03}.
However, metal-rich GCs in giant ETGs are generally thought to
have formed in high-pressure environments similar to those seen in vigorously
star-forming galaxies at high redshift \citep[e.g.,][]{ashzep92,krui15} where
SFRs exceed $10^2 \; M_{\odot} \; {\rm yr}^{-1}$ and turbulent 
velocities are of order $\sigma = 10 - 100$
\kms\ \citep{law+09,law+12,genz+10,tacc+13}. 
In such harsh environments, the crossing time of a (proto)cluster with a
typical half-mass radius $r_{\rm h} \sim 1$ pc is only of order
$\tau_{\rm cross} = r_{\rm h}/\sigma = 10^4 - 10^5$ yr, which is 1\,--\,2
orders of magnitude shorter than in the solar neighborhood. Fokker--Planck
modeling of GCs suggests that the evaporation time of a multi-mass cluster is
of order $\tau_{\rm ev} \approx 10 \; \tau_{\rm rel}$ where $\tau_{\rm rel}$
is the half-mass relaxation time of the cluster
\citep[e.g.,][]{leegoo95}. Since $\tau_{\rm rel}/\tau_{\rm cross} \sim 0.1\,N/\ln N$
where $N$ is the number of stars in the cluster
\citep{bintre87}, it follows that for a cluster to survive evaporation over a
star formation time scale of $10^7$ yr in such high-pressure environments, the
requirement is $0.1\,N/\ln N \ga 10 - 100$, corresponding to
$N \ga 1500 - 5000$ cluster stars. With this in mind, we perform calculations
of $f_{\rm GC}$ with  $10^2 \leq \cM_{0,\,{\rm min}}/M_{\odot} \leq 10^4$. 
Results are shown in Figure~\ref{f:Mfracplot}. Panels (a)\,--\,(c) plot
$f_{\rm GC}$ versus 
$\cM_{\rm limit}$, while panels (d)\,--\,(f) plot $\Delta f_{\rm GC}$ versus
$\cM_{\rm limit}$, where
$\Delta f_{\rm GC} \equiv f_{\rm GC} \,(\cM_{\rm limit}, \cM_{\rm c},
\mu_{\rm ev})- f_{\rm GC} \,(\cM_{\rm limit}, 10^6\;M_{\odot},
10^4\;M_{\odot}\;{\rm Gyr}^{-1})$,
i.e., the value of $f_{\rm GC}$ relative to that for
$\cM_{\rm c} = 10^6\;M_{\odot}$ and
$\mu_{\rm ev} = 10^4\;M_{\odot}\;{\rm Gyr}^{-1}$.

Note that the range of $\Delta f_{\rm GC}$ implied by the hypothesis that the
range of \FUVmV\ seen among ETGs is produced by (now largely dissolved) massive 
metal-rich GCs in the inner regions can indeed be reproduced by dynamical
evolution of such GCs under quite reasonable conditions. 
Specifically, we find that the range $\Delta f_{\rm GC} \la 0.23$ (cf.\ 
Section~\ref{sub:ener}.1) 
is covered 
if $\cM_{\rm c}$ is in the range $10^6 \la \cM_{\rm c}/M_{\odot} \la 10^7$.
Encouragingly, this range of $\cM_{\rm c}$ implied by our hypothesis is
\emph{entirely consistent with results of fits of Equation~(\ref{eq:finMF}) (for
$\gamma = 1$) to observed GC luminosity functions (GCLFs) of giant ETGs}, for
which $\cM_{\rm c}$ is found to increase with ETG luminosity and hence
presumably also ETG mass \citep{jord+07,john+17}\footnote{Note that the values
  for $\cM_{\rm c}$ in \citet{jord+07} do not account for mass loss due to
  stellar evolution; the ones in \citet{john+17} do.}.  Generally,
Figure~\ref{f:Mfracplot} shows that $f_{\rm GC}$ for ETGs with the
weakest UV upturns is reached for low values for $\cM_{\rm c}$
and/or $\mu_{\rm ev}$, while $f_{\rm GC}$ values for strong UV upturns
require higher values for $\cM_{\rm c}$ and/or $\mu_{\rm ev}$.  

Summarizing this section, we find that the range of observed \FUVmV\ colors in
the inner regions of nearby ETGs is consistent with our hypothesis
that the range in \FUVmV\ seen among ETGs is produced by He-rich EHB stars
associated with massive ($\log\,(\cM_0/M_{\odot}) \ga 5.5$) metal-rich GCs,
most of which have dissolved after $\sim$\,12 Gyr of dynamical evolution. The
\FUVmV\ colors of ETGs with the weakest UV upturns and low values of $S_N$ are
consistent with GCs that were formed in environments featuring relatively low
SFRs, associated with relatively low characteristic truncation
masses ($\cM_{\rm c} \approx 10^6\;M_{\odot}$), whereas the GCs in ETGs with
the strongest UV upturns and high $S_N$ values were likely formed in vigorously
star-forming environments where $\cM_{\rm c} \approx 10^7\;M_{\odot}$. This
result is supported by the recent numerical simulations of GC
formation at high redshift by \citet{li+17}, who found a
correlation between $\cM_{\rm c}$ and SFRs such
that $\cM_{\rm c} \propto {\rm SFR}^{1.6}$. 

\subsubsection{The Connection between $\cM_{\rm c}$ and $S_N$}

While the previous section clarified that the range of $f_{\rm GC}$ implied by 
the observed range of \FUVmV\ among ETGs in the context of our hypothesis 
can be explained by a range in $\cM_{\rm c}$ among ETGs similar to that
observed, it does not directly explain how the observed anticorrelation
between \FUVmV\ and $S_N$ seen in Figures~\ref{f:FUVmVfig1} and
\ref{f:FUVmVfig2} fits in with this scenario. This is addressed in this
section.  

Following the original definition of the specific frequency of GCs by 
\citet{harvdb81}, $S_N$ values are determined by parameterizing the observed
completeness-corrected GCLF as a Gaussian in magnitude units. After fitting 
the turnover magnitude $M_{\rm TO}$ of the Gaussian, the total number of GCs is
determined by counting the number of GCs brighter than $M_{\rm TO}$ and doubling
that value. (The reason behind this methodology is to avoid
incompleteness-related issues at faint magnitudes.)

\begin{figure*}
\centerline{\includegraphics[width=13cm]{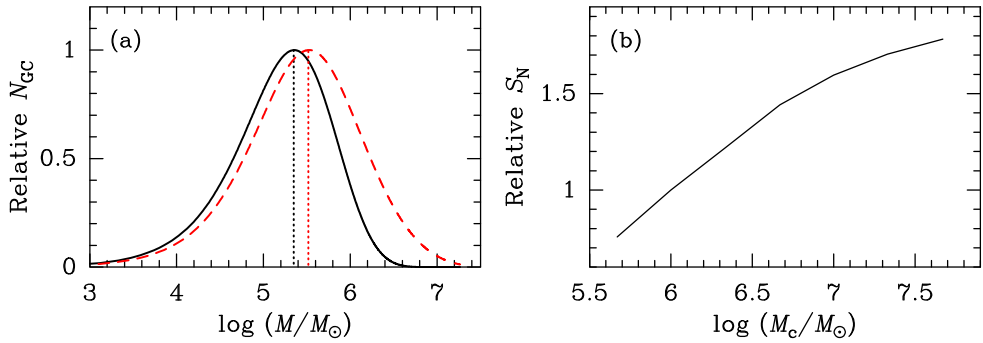}}
\caption{\emph{Panel (a)}: GCMFs (i.e., Equation \ref{eq:finMF}) at $t$ = 12 Gyr
  for $\mu_{\rm ev} = 3\times10^4\; M_{\odot} \; {\rm Gyr}^{-1}$. The black solid
  line is for $\cM_{\rm c} = 10^6\; M_{\odot}$ while the red dashed line is for
  $\cM_{\rm c} = 10^7\; M_{\odot}$. For reference, the two vertical dashed
  lines indicate the values of the turnover mass
  for the two GCMFs. Note the significant difference in GCMF width between the
  two values of $\cM_{\rm c}$. 
  \emph{Panel (b)}: $S_N$ versus $\cM_{\rm c}$ at $t$ = 12 Gyr
  for $\mu_{\rm ev} = 3\times10^4\; M_{\odot} \; {\rm Gyr}^{-1}$. Note that a range
  of $\cM_{\rm c}$ suggested by the discussion in Section~\ref{sub:ener}.2
  causes a range in $S_N$ of a factor $\sim$\,1.7 at a given evaporation rate. 
  }
\label{f:Mc_figure}
\end{figure*}

In the presence of a range of $\cM_{\rm c}$ among GCMFs of ETGs where 
$\cM_{\rm c}$ increases with galaxy luminosity and mass, it is instructive to
test the sensitivity of $S_N$ to $\cM_{\rm c}$, because the latter affects the
width of the GCLF at old age. This is illustrated in panel (a) of
Figure~\ref{f:Mc_figure}, while we plot $S_N$ as a function of $\log\,\cM_{\rm
  c}$ at an age of 12 Gyr in panel (b) of Figure~\ref{f:Mc_figure}. These 
values of $S_N$ were derived by evaluating Equation~(\ref{eq:finMF}) for $\gamma$ =
1, $\mu_{\rm ev} = 3 \times 10^4\; M_{\odot} \; {\rm Gyr}^{-1}$, $t$ = 12 Gyr, and
a range of $\log\,\cM_{\rm c}$. Note that the aforementioned estimated range 
$10^6 \la \cM_{\rm c}/M_{\odot} \la 10^7$ for ETGs with UV upturns
implies a range in $S_N$ of a factor $\sim$\,1.7, \emph{which is similar to the range
in $S_{N,\,z,\,{\rm red}}$ actually found among the galaxies in our sample}  
(see Figures~\ref{f:FUVmVfig1}(f) and \ref{f:FUVmVfig2}).

We thus arrive at a picture where the observed ranges of both \FUVmV\ and
$S_{N,\,z,\,{\rm red}}$ can be largely explained by the observed range in
$\cM_{\rm c}$, which in turn can be explained by a range of SFR \citep[or SFR surface
density, see also][]{john+17} occurring in the progenitors of the present-day
ETGs, with the more massive ETGs containing higher mass fractions from more
massive protogalaxies, which formed their stars earlier and with
higher SFRs than the less massive ones.

\subsection{Metallicities of Metal-rich GCs and the Underlying Field Population}
\label{sub:ZH}

As to the actual metallicities of the metal-rich GCs in the FUV-emitting
(inner) regions of ETGs, it has been suggested that they are typically lower
than that indicated by the underlying diffuse light of the parent galaxy
\citep[e.g.,][]{peng+06,sohn+06}, thus casting some apparent doubt on whether
these GCs might be able to explain the UV upturns in ETGs through the scenario
advocated here. However, we emphasize that metallicity data on GCs in the inner
regions of ETGs (including M87) are virtually always based on optical photometry
rather than spectroscopy.\footnote{This is due to signal-to-noise constraints
  imposed by the high background level in the central regions of ETGs.} 
As shown by \citet{goukru13}, the differences in optical colors
between giant ETGs and their metal-rich GCs can be explained by differences in
the stellar MFs. Specifically, several recent spectroscopic
studies of giant ETGs have established evidence for bottom-heavy stellar MFs,
with power-law slopes $\alpha$ in the sub-solar mass range that are at least
as steep as that of the initial mass function (IMF) of \citet{salp55}, i.e.,
$-3.0 \la \alpha \la -2.3$ in $dN/d\cM \propto \cM^{\alpha}$ 
\citep[e.g.,][]{vancon10,convan12,smit+12b,laba+13,mcde+14,spin+14,vand+17,sarz+18}.
For GCs that were born with such bottom-heavy IMFs, \citet{goukru13} showed
that 12 Gyr of dynamical evolution causes the stellar MF of the average
surviving (massive) GC to be similar to the canonical \citet{krou01} IMF, thereby
causing their optical colors to become bluer than the underlying field
population by amounts similar to those observed \citep[see
  also][]{goukru14}. As such, the bluer colors of metal-rich GCs relative to
their parent galaxies do not necessarily indicate differences in
metallicity, which would indeed be unexpected under the common assumption that
surviving GCs represent the high-mass end of the star formation processes that
also created the field population \citep{elmefr97}.  Furthermore, in the few
cases where high-quality spectroscopic Lick index data are available for both 
ETGs and their constituent GCs at comparable galactocentric distances, the
metal-rich GCs have the same ages and metallicities as their parent ETGs to
within small uncertainties (see \citealt{goukru13} and references therein). 
This is also the case for M87, as detailed in the Appendix.

\section{Implications of the Proposed Scenario}
\label{s:implications}

\subsection{UV Upturn: Metal-poor and/or Metal-rich Populations?}
\label{sub:MPvsMR}

Traditionally, the correlation of UV upturn strength with the Mg$_2$ index
among ETGs has always been recognized as somewhat surprising, in the sense 
that the Galactic GCs with the strongest FUV fluxes are typically metal-poor 
\citep[see, e.g.,][]{dorm+95,ocon99}. One early scenario that was proposed to
resolve this paradox postulated that the UV upturn was mainly produced by a
minority ($\la$\,20\%) population of  metal-poor stars \citep{parlee97},
although this scenario required ages of $\ga 16$ Gyr for the
metal-poor component. More recently, the fact that the metal-poor GCs
in M87 are on average brighter in the FUV than the metal-rich GCs (see 
\citealt{sohn+06}) may have prompted \citet{chun+11} to
construct a model in which the fraction of helium-enhanced stars decreases with 
increasing metallicity, leading them to conclude that the bulk of the
FUV flux in ETGs is produced by metal-poor (and He-rich) stars.
While the model of \citet{chun+11} produces a very good fit to the observed
FUV upturn and optical SEDs of ETGs,
it overproduces flux in the NUV (2000\,--\,3000 \AA) region (see their
Figure~3), which is consistent with a  lack of metallicity-dependent
line blanketing in the NUV in their model \citep{dorm+95,tant+96}. 
Furthermore, the metallicity distribution of the FUV-bright
population in the model of \citet{chun+11} seems inconsistent with the observed
anticorrelations between radial gradients of \FUVmV\ and metallicity within
ETGs \citep[both in Mg$b$ and \ZH,][]{jeon+12}, which are in the sense
that UV upturn strength and metallicity both decrease with increasing
galactocentric radius, while age does not. This suggests a physical
association between UV upturn strength and metallicity.

\begin{figure*}
\centerline{\includegraphics[width=13.5cm]{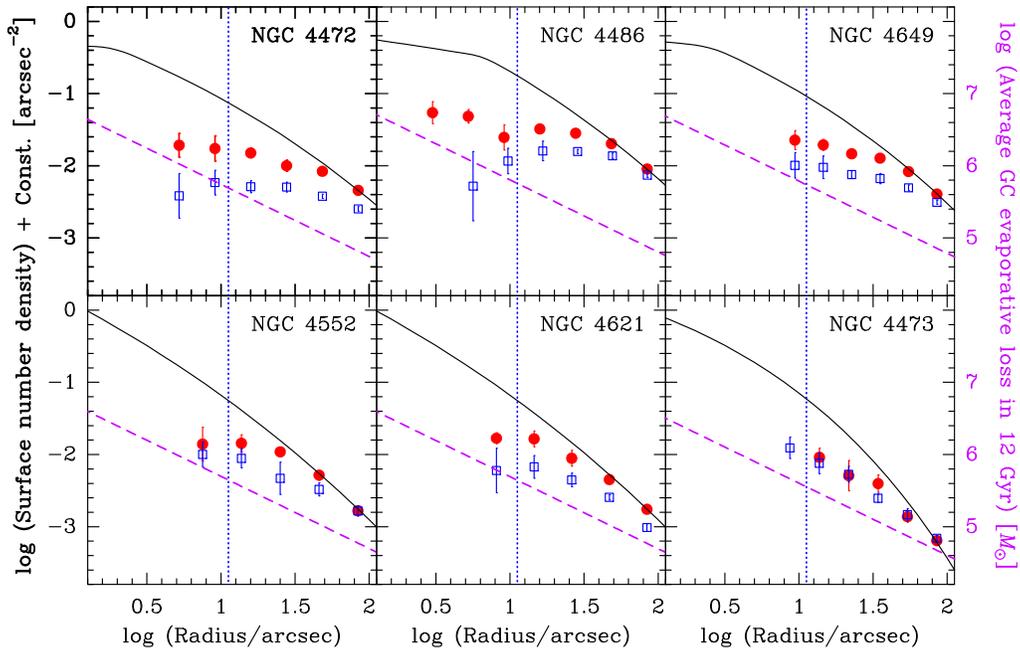}}
\caption{A copy of Figure~\ref{f:radprof} to which we added radial surface
  number density profiles of metal-poor GCs in each galaxy, shown with open
  blue squares. See discussion in Section~\ref{sub:MPvsMR}.  
  }
\label{f:radprof_blue_red}
\end{figure*}

\begin{figure*}
\centerline{\includegraphics[width=12.5cm]{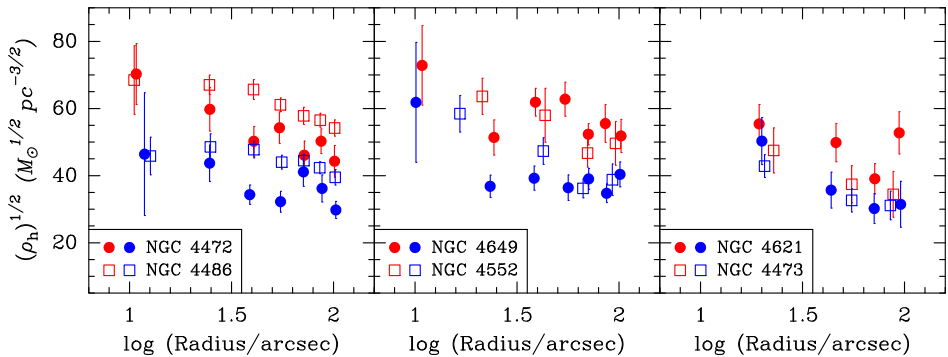}}
\caption{Radial profiles of the running average value of $\rho_{\rm h}^{1/2}$ 
  for the GC systems in the 6 ACSVCS galaxies shown in Figures~\ref{f:radprof} and
  \ref{f:radprof_blue_red}. Blue symbols correspond to metal-poor GCs while red
  symbols correspond to metal-rich GCs. See legend for the associations between
  symbol types and galaxies. See discussion in Section~\ref{sub:MPvsMR}.   
  }
\label{f:muplot}
\end{figure*}

In the context of the scenario described here, this physical association is
implied by the differences in radial distribution and structural parameters
between the metal-rich and the metal-poor GCs. The radial distributions of the
two GC subpopulations in the central $\sim$\,8 kpc of giant ETGs in the ACSVCS are
shown in Figure~\ref{f:radprof_blue_red}, which is a copy of
Figure~\ref{f:radprof} to which we added data points for the metal-poor
GCs. Note that the surface number densities of metal-poor GCs in the inner
regions are factors of 2\,--\,5 \emph{lower} than those of the metal-rich GCs
in ETGs with the strongest UV upturns such as M87 and NGC 4649. It is important to
realize that this difference between the metal-poor and metal-rich GC systems
in these galaxies is \emph{not} due to differences in GC evaporation
rates. First, as already mentioned in Section~\ref{sub:GCdiss}.1,  
the radial distribution of metal-rich GCs on larger radial scales (10\,--\,100
kpc) is consistent with that of the galaxy light whereas that of the metal-poor GCs
is significantly more shallow. Secondly, the observed structural parameters of
the GCs in these galaxies indicate that the evaporation rates of the metal-rich
GCs are actually \emph{higher} than those of the metal-poor GCs at a given 
projected $R_{\rm gal}$. As shown by \citet{goufal16}, the evaporation rate
$\mu_{\rm ev}$ of tidally limited GCs scales with their mass density as
$\mu_{\rm ev} \propto \rho_{\rm  h}^{1/2}$, where
$\rho_{\rm h} = 3 \cM / (8 \pi r_{\rm h}^3)$ is the mean
density of a GC within its half-mass radius (see also \citealt{mclfal08}).
Figure~\ref{f:muplot} depicts the rolling mean value of $\rho_{\rm h}^{1/2}$
versus $R_{\rm gal}$ for the metal-poor and metal-rich GCs in the six ACSVCS
galaxies also shown in Figure~\ref{f:radprof_blue_red}.
Note that $\rho_{\rm h}^{1/2}$ is systematically higher in metal-rich GCs than
in metal-poor GCs. Statistically, the evaporation rate as measured by
$\rho_{\rm  h}^{1/2}$ in metal-rich GCs is larger than that in metal-poor ones
by a factor $1.32 \pm 0.12$  at a given $R_{\rm gal}$ among these six galaxies. 

Even though the massive metal-poor GCs in M87 are brighter in the FUV than
their metal-rich counterparts by a factor $\sim$\,2 on average 
(see Figure~\ref{f:FUVmV_vs_mass}), the metal-rich GCs have significantly
higher surface number densities 
\emph{as well as} higher evaporation rates than the metal-poor GCs in the
central regions of the ETGs with the strongest UV upturns. It follows that the
FUV emission is due mainly to metal-rich GCs in our scenario. This is
consistent with the observed anticorrelations between radial gradients of
\FUVmV\ and metallicity within ETGs as mentioned above.

\subsection{Light-element Abundance Ratios in ETGs}
\label{sub:X_Fe}

If the scenario proposed here is correct, an important implication would be that
the He-enhanced populations responsible for the UV upturn in ETGs would show
light-element abundance ratios similar to those observed in massive GCs, i.e.,
enhancements in [Na/Fe] and [N/Fe] accompanied by \citep[likely more
  metallicity-dependent, see][]{vent+13} depletions of [C/Fe] and [O/Fe].  

Specifically, the amplitude of light-element abundance variations in GCs is
known to scale with GC mass (see Section~\ref{s:intro}), so that the largest
such variations would be found in regions where the mass fraction of stars
originating from massive GCs is highest. In the current scenario, that occurs
where $\cM_{\rm c}$ and $\mu_{\rm ev}$ are highest, i.e., the same regions
where the UV upturn is strongest---the inner regions of ETGs, while the
amplitude of these abundance variations would decrease toward the outer
regions of ETGs. Interestingly, some recent spectroscopic studies do show
evidence for this. Using stacked SDSS spectra of ETGs, \citet{schi07} found a
strong correlation between [N/Fe] and galaxy luminosity, while \citet{vand+17}
used deep spatially resolved spectroscopy of six ETGs and found that [Na/Fe]
increases toward the galaxy centers within $R_{\rm gal} \la R_{\rm eff}/2$
\citep[see also][]{sarz+18}, 
while [O/Fe] decreases. These results are just as expected in the current
scenario where the He enhancement is due to dissolution of massive GCs in the
central regions. In this respect, a clear prediction from the current scenario
is that future spectroscopic studies will also find radial gradients in [N/Fe]
that increase toward galaxy centers in massive ETGs, and that such radial
gradients have the largest amplitudes in ETGs with the strongest UV
upturns.

\subsection{Interpretation of Correlations between UV Upturn and Galaxy Properties}
\label{sub:interpret}

If indeed the range of \FUVmV\ among ETGs is mainly caused by He-rich EHB stars
that were formed in massive GCs, among which most were subsequently dissolved in the
inner regions of giant ETGs, can we understand the previously known relations
between \FUVmV\ and galaxy properties such as those shown in panels
(a)\,--\,(e) of Figure~\ref{f:FUVmVfig1}? 

First of all, the anticorrelation between \FUVmV\ and \aFe\ can be
understood, at least qualitatively, by the results of \citet{puzi+06} who found
that massive GCs in ETGs typically have very high \aFe\ values, higher than
those of the diffuse light of their parent galaxies. Hence, the higher the
values of $\cM_{\rm c}$ and/or average  $\mu_{\rm ev}$ for a given ETG, the
higher the resulting \aFe, as observed. 
Similarly, the anticorrelation between \FUVmV\ and central velocity dispersion
arises in this scenario due to the strong correlation between ETG luminosity
(and thus likely mass as well) and the value of $\cM_{\rm c}$ of its GC system
(see Section~\ref{sub:ener}.3 and \citealt{john+17}), in conjunction with the
fact that $\mu_{\rm ev} \propto \sigma$ (see Equation \ref{eq:mu}).
This would also indirectly cause the anticorrelation between \FUVmV\ and \ZH,
given the well-known mass-metallicity relation among ETGs \citep[e.g.,][]{ferr+06}. 
The anticorrelation between \FUVmV\ and age would arise in part due to the
age dependence of the production rate of EHB stars due to stellar evolution,
and in part because the more massive ETGs (with higher values of $\cM_{\rm c}$
and $\mu_{\rm ev}$) typically have older SSP ages. Finally, the strong 
\citet{burs+88} anticorrelation between \FUVmV\ and Mg$_2$ can be explained by
the sensitivities of the Mg$_2$ index to age, \ZH, and \aFe, all of which
anticorrelate with \FUVmV\ \citep[see also][]{jeon+12}. 

\citet{zari+15} recently reported a correlation between the strength of the UV
upturn in ETGs and the stellar mass-to-light ratio inferred from SED fitting,
suggesting that differences in the low-mass end of the stellar IMF are related
to the nature of the EHB stars responsible for the FUV flux in ETGs. In the
context of the scenario on the nature of the UV upturn presented here, the
relation found by \citeauthor{zari+15}\ arises for the same reasons as that of the
relation between $\cM_{\rm c}$ and ETG luminosity or mass, i.e., 
a scenario in which the central regions in massive ETGs contain
relatively high mass fractions from massive galaxy building blocks, 
which formed their stars earlier and with higher SFR surface densities
than less massive ones. 
Those very high SFR surface densities are also thought to cause the steep
stellar IMFs at sub-solar masses found in the central regions of massive ETGs,
specifically through strong turbulence (high Mach numbers), which causes cloud
fragmentation to occur at relatively small scales \citep{hopk13}. 

\section{Summary and Conclusions}
\label{s:conc}

Prompted by the recent finding of FUV-bright massive GCs in M87, the central
dominant galaxy in the Virgo cluster of galaxies
\citep{sohn+06,kavi+07},  we investigate the idea that there is a
physical connection between the UV upturn in ETGs and He-enhanced stellar
populations in massive GCs.  
We study the dependencies of the strength of the UV upturn in ETGs
on the GC specific frequency $S_N$ and other galaxy properties, mainly using
results from the literature. We find that \FUVmV\ anticorrelates strongly with
$S_{N,\,{\rm red}}$, the specific frequency of red (metal-rich) GCs in
ETGs. This anticorrelation appears to be causal, in that ETGs with high values
of $S_{N,\,{\rm red}}$ consistently lie ``above'' linear fits to the previously
known anticorrelations between \FUVmV\ and Mg$_2$, age, \ZH, and central
velocity dispersion $\sigma$.

Guided by the observed depletion of surface number densities of metal-rich
GCs in the inner regions of massive ETGs, which is where the UV upturn is known 
to occur, we explore the hypothesis that the UV upturn is produced mainly by
He-enhanced populations formed within massive (mainly metal-rich) GCs that are
subsequently disrupted by dynamical evolution in the strong tidal fields in the
inner regions of ETGs during the $\ga 10$ Gyr of their lifespan, using the
observed FUV luminosities seen in the surviving massive GCs in M87 as proxies. 

Adopting a \citet{sche76} function parameterization of GCMFs in conjunction with
simulations of dynamical evolution of GCs, we find that the ranges of observed
\FUVmV\ colors and FUV luminosities among ETGs are entirely consistent with our
hypothesis if the initial masses of GCs responsible for the bulk of the FUV
output were $\cM_0 \ga 3\times10^5 \; M_{\odot}$. (This value for $\cM_0$  is
formally an underestimate, since it does not account for the effects of
rapid mass loss mechanisms during the first $\approx 10^8$ yr after GC
formation.) This lower limit of the initial mass of GCs responsible for 
the UV upturn is consistent with the masses of the FUV-bright GCs currently
seen in M87. Specifically, we find that the \FUVmV\ colors of the ETGs with the
weakest UV upturns and lowest values of $S_{N,\,{\rm red}}$ are consistent with
GCs that were formed in environments featuring relatively low characteristic
Schechter truncation masses ($\cM_{\rm c} \approx 10^6\;M_{\odot}$), likely
associated with regions with relatively low SFR surface densities. Conversely, 
the \FUVmV\ colors of the ETGs with the strongest UV upturns and highest
$S_{N,\,{\rm red}}$ values are consistent with GC systems with
$\cM_{\rm c} \approx 10^7\;M_{\odot}$, likely having formed in vigorously
star-forming environments. Importantly, the values of $\cM_{\rm c}$ necessary
to explain the range of \FUVmV\ seen among ETGs are consistent with the values
of $\cM_{\rm c}$ found from evolved Schechter function fits of the GC
luminosity functions of those ETGs. Furthermore, we find that this range of
$\cM_{\rm c}$ found among ETGs also explains the correlation between
\FUVmV\ and $S_{N,\,{\rm red}}$ in that GCMFs with larger values of
$\cM_{\rm c}$ have wider GCMFs, which translates to larger values of $S_N$ due
to the way $S_N$ is defined.

If the scenario proposed here is correct, an important implication would be that
the He-enhanced populations responsible for the UV upturn in ETGs should show
light-element abundance ratios similar to those observed within massive GCs,
i.e., enhancements of [Na/Fe] and [N/Fe] accompanied by depletions of [C/Fe] and
[O/Fe]. Encouragingly, a recent spectroscopic study of massive ETGs by
\citet{vand+17} does show evidence for radial gradients of [Na/Fe] and [O/Fe]
in the sense predicted by this scenario. As such, we predict that future
studies will also find radial gradients in [N/Fe] that increase toward galaxy
centers in massive ETGs, and that such radial gradients have the largest
amplitudes in ETGs that have the strongest UV upturns.  

Our findings suggest that the nature of the UV upturn in ETGs and the
variation of its strength among ETGs are causally related to that of
helium-rich populations in massive GCs, rather than intrinsic properties of
field stars in massive galactic spheroids.
The observed ranges of both \FUVmV\ and $S_{N,\,{\rm red}}$ among ETGs can be
explained by the observed range in $\cM_{\rm c}$ of their GC systems, which in
turn can be explained by a range of SFR surface density occurring in the
progenitors of the present-day ETGs, with the more massive ETGs containing
higher mass fractions from more massive protogalaxies, which formed their stars
earlier and with higher SFR surface densities than the less massive ones.

\acknowledgments
I thank the anonymous referee for a very thoughtful report with
relevant and helpful suggestions. 
I also acknowledge useful discussions with Thomas Puzia, Tom Brown, Marcio
Catel\'an, and Diederik Kruijssen.  
This research made use of the HyperLeda database
(http://leda.univ-lyon1.fr). 
This research has made use of the NASA/IPAC Extragalactic
Database (NED), which is operated by the Jet Propulsion Laboratory, California
Institute of Technology, under contract with the National Aeronautics and Space
Administration. This paper is based in part on observations made with the NASA
Galaxy Evolution Explorer (GALEX). GALEX was operated for NASA by the California
Institute of Technology under NASA contract NAS5-98034. 

%






\appendix

\section{A Comparison of Metallicities of Metal-rich GCs in M87
  with the Underlying Field Population}
\label{s:M87_spectra}

\begin{figure}[tbph]
\centerline{\includegraphics[width=8.3cm]{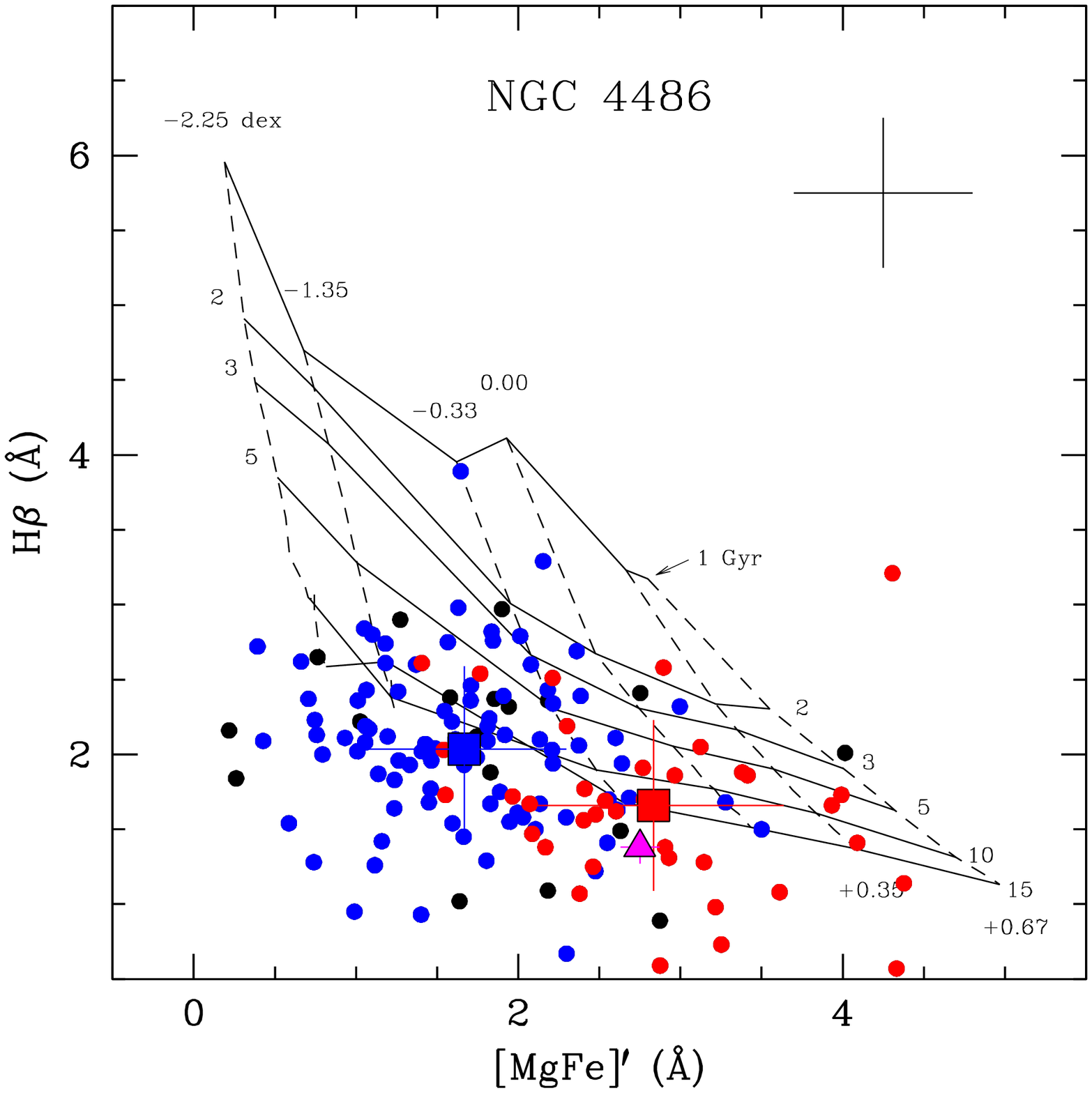}}
\caption{H$\beta$ versus [MgFe]$'$ for GCs in NGC~4486 (= M87) from the data of 
  \citet{cohe+98}. Small blue and red circles indicate GCs designated 
  metal-poor and metal-rich, respectively, by means of $U\!-\!R_J$ colors from
  \citet{stro+81}. Large filled blue and red squares represent weighted
  average values of H$\beta$ and [MgFe]$'$ of 
  the metal-poor and metal-rich GCs, respectively.
  GCs. SSP models of \citet{thom+03} are overplotted for [$\alpha$/Fe] =
  +0.3. Dashed lines indicate \ZH\ values of $-$2.25, $-$1.35, $-$0.33, 0.00,
  0.35, and 0.67 dex.
  Solid lines indicate ages of 1, 2, 3, 5, 10, and 15 Gyr.
  For comparison, the magenta triangle represents H$\beta$
  and [MgFe]$'$ for the diffuse light of NGC~4486 at $\Rgal = 40''$. See
  discussion in Appendix A.  
\label{f:M87specplot}
}
\end{figure}

For the specific case of GCs in M87, the available spectroscopic data in
the literature \citep{cohe+98} are, unfortunately, strongly dominated by GCs
at large galactocentric distances, and thus by metal-poor GCs. Furthermore, the
S/N ratio of a significant fraction of the GC spectra of \citet{cohe+98} are
not high enough to derive ages and metallicities at the $\sim$\,50\% accuracy
level (S/N $\sim$\,30 per \AA, see \citealt{puzi+05}), thus providing only
marginal constraints on the age and metallicity of these GCs.
To reduce the resulting uncertainty on the assignment of GCs targeted by
\citet{cohe+98} as ``metal-poor'' versus ``metal-rich,'' we use the $U\!-\!R_J$
(where $R_J$ is Johnson $R$) colors from \citet{stro+81}, whose GC target list
was used by \citet{cohe+98}. The selection of metal-poor (or metal-rich) GCs was
made by selecting GCs bluer (or redder) than the $U\!-\!R_J$ color associated
with the metallicity corresponding to the dip in $g\!-\!z$ between the blue and
red peaks in the bimodal color distribution in the high-quality ACSVCS
photometry of M87 \citep[$g\!-\!z = 1.20$,][]{peng+06}. In this context, we use
BC03 model SEDs for an age of 12 Gyr and a \citet{chab03} IMF, and derive model
$U\!-\!R_J$ colors by using the {\sc synphot} package within IRAF/STSDAS. The
resulting $U\!-\!R_J$ color to discriminate metal-poor from metal-rich GCs in
M87 is 1.75.  
To evaluate mean metallicities for the metal-poor and metal-rich GCs, we use the
indices H$\beta$ and [MgFe]$'$,
respectively, and compare them to predictions of the SSP models of
\citet{thom+03}. 
Figure~\ref{f:M87specplot} shows H$\beta$ versus [MgFe]$'$ for the GCs in M87
from \citet{cohe+98}.
The metal-poor and metal-rich GCs selected as such by means of their $U\!-\!R_J$
colors from \citet{stro+81} are shown with blue and red circles, respectively. 
Since the uncertainties are significant for the individual GCs, we also
indicate inverse-variance-weighted average indices for the metal-poor and
metal-rich GCs selected as mentioned above (see large blue and red squares in 
Figure~\ref{f:M87specplot}).  
Note that \ZH\ $\simeq -0.25$ for the weighted average metal-rich GC, for which 
$2'' \leq \Rgal \leq 390''$ with a median of 181$''$. 
For comparison, we overplot H$\beta$ and [MgFe]$'$ for the diffuse light of
M87 at $\Rgal \sim 40''$, the outermost data point from \citet{davi+93} and
\citet{sarz+18}. Note that the age and metallicity of the latter are consistent
with those of the average metal-rich GC in M87 to within 1\,$\sigma$. We
conclude that the available spectroscopic observations of M87 and its GCs are
consistent with previous evidence for other ETGs (see \citealt{goukru13} and
references therein) in that the mean metallicities of metal-rich GCs and the
underlying field stars in giant ETGs are consistent with each other. 

\end{document}